\pdfoutput=1
\documentclass[a4paper,american,floatfix,pdftex,superscriptaddress,twoside,%
aps,prl,%
reprint
]{revtex4-2}%
\usepackage[T1]{fontenc}
\usepackage{multirow}
\usepackage{graphicx}%
\usepackage[utf8]{inputenc}
\usepackage{hyperref, hypernat}
\usepackage[ams,todos]{CMImacros}
\usepackage[boxed]{algorithm2e}
\graphicspath{{./figures/}}
\usepackage{tabularx}
\usepackage{longtable}
\usepackage{booktabs}
\usepackage{adjustbox}
\newcommand{\cfeldesy}{\affiliation{Center for Free-Electron Laser Science CFEL, Deutsches
      Elektronen-Synchrotron DESY, Notkestr. 85, 22607 Hamburg, Germany}}%
\newcommand{\westlake}{\affiliation{Zhejiang Key Laboratory of 3D Micro/Nano Fabrication and Characterization, Westlake Institute for Optoelectronics, Fuyang, Hangzhou, Zhejiang 311421, China}}%

\newcommand{\ayemail}{\email[Email:~]{andrey.yachmenev@robochimps.com}}%
\newcommand{\gyemail}{\email[Email:~]{yangguang@wioe.westlake.edu.cn}}%

\SetAlCapSkip{\abovecaptionskip}

\begin{document}

\title{Nuclear spin symmetry-breaking and spin polarization in rotational energy level clusters}%
\author{Andrey Yachmenev}\ayemail\cfeldesy%
\author{Guang Yang}\gyemail\westlake
\date{\today}

\begin{abstract} 
We present the first quantum mechanical study of hyperfine effects in the rotational cluster states of a symmetric triatomic molecule H$_2$S.
Rotational clusters arise from spontaneous symmetry breaking induced by high-angular-momentum rotational motions in certain rigid molecules, resulting in dynamic enantiomorphism driven by kinetic distortion effects.
Hyperfine interactions in the cluster states lead to collision-free breaking of nuclear spin symmetry, with the magnitude of nuclear spin \emph{ortho}-\emph{para} mixing significantly exceeding that in other states with same or lower angular momentum.
The \emph{ortho}-\emph{para} mixing induces nuclear spin polarization in the laboratory frame and gives rise to two sets of enantiomers, that have different energies and oppositely oriented nuclear spin projections.
Although hyperfine interactions preserve parity, they lift the degeneracy of opposite-parity cluster states.
This phenomenon, previously observed experimentally, is explained as a result of tunneling between rotating enantiomers, facilitated by the Pauli exclusion principle.
\end{abstract}

\maketitle

Chirality, in its traditional sense, refers to the time-invariant property of certain molecules to exist in two nearly energetically equivalent configurations, known as enantiomers, which are non-superimposable mirror images of each other.
This means, enantiomers cannot be brought into coincidence through spatial rotations or translations.
The concept of chirality has been broadened to include time-noninvariant enantiomorphism, named ``false chirality'', which extends the original notion by adding time reversal as the operation that interconverts the enantiomers~\cite{Zocher_PNAS39_1953, Barron_ChemSocRev15_1986, Barron_MagnetoChem6_2020}.
Falsely chiral influences, such as, for example, the interaction of chiral molecules with magnetic fields~\cite{BanerjeeGhosh_Science360_2018, Naaman_AnnuRevBiophys51_2022, Bloom_ChemRev124_2024} or chiral electric fields~\cite{Yachmenev_PRL123_2019, Mayer_PRL129_2022, Mayer_NatPhot18_2024}, can distinguish between enantiomers, however only in processes that are either prevented from reaching equilibrium or have not yet reached it~\cite{Barron_JACS108_1986, Barron_CPL135_1987}.

Much like the potential energy surface of a statically chiral molecule has two equivalent minima corresponding to molecular structures related by parity inversion, kinetic energy can similarly localize molecular motions into two parity-related configurations, making a statically achiral molecule dynamically chiral.
This form of dynamic chirality can arise from kinetic effects such as centrifugal distortion and Coriolis coupling, that is, under non-equilibrium conditions, and falls into the category of false chirality.

A notable example is the rotational energy level clustering effect~\cite{Harter_JCP80_1984, Pavlichenkov_PhysRep226_1993,Jensen_WIRECompMolSci2_2012}, where the rovibrational motions of a non-chiral molecule become localized around several stable axes of rotation in the molecular frame.
These axes are separated by high kinetic energy barriers, leading to a symmetry-breaking of the molecule~\cite{Bunker_JMolSpec228_2004}.
Other examples of kinetic effects include the dynamical symmetry breaking in structural transitions of atomic clusters~\cite{Oka_JCP142_2015}, chirality-dependent coupling of molecular rotation and translation in mixtures~\cite{Evans_PRL55_1985}, and the transfer of macroscopic chirality in fluid and mass transport in helical devices~\cite{Sevim_NatCommun13_2022}.

The phenomenon of rotational energy level clustering occurs in molecules exhibiting local mode behavior~\cite{Jensen_MP98_2000}, where molecular bonds are nearly orthogonal to each other such that the rotation of the molecule remains largely unaffected by Coriolis-type interactions with vibrational modes~\cite{Jensen_WIRECompMolSci2_2012}.
This effect has been investigated extensively in quantum and semiclassical calculations for symmetric triatomic $X$H$_2$-type molecules (e.g., $X=$~S~\cite{Kozin_JMolSpec163_1994,Owens_JPCL9_2018}, P~\cite{Yurchenko_PCCP7_2005, Owens_PRL121_2018}, Si~\cite{Clark_JQSRT246_2020}, Po~\cite{Gomez_JMolSpec186_1997}, Te~\cite{Gomez_JMolSpec185_1997,Jensen_ChemPhys190_1995}, Se~\cite{Kozin_JMolSpec161_1993}) and tetratomic $X$H$_3$-type molecules (e.g., $X=$ P, Bi, Sb)~\cite{Yurchenko_JMolSpec240_2006}, their deuterated isotopologues~\cite{Yurchenko_JMolSpec256_2009}, as well as for SO$_3$~\cite{Underwood_JCP140_2014} and PF$_3$~\cite{Mant_MP118_2019}.
Experimentally, rotational clusters were observed in H$_2$Se~\cite{Kozin_JMolSpec152_1992, Flaud_JMolSpec172_1995} and dimethylsulfoxide~\cite{Cuisset_PRL109_2012, Cuisset_JCP138_2013}.
Additionally, several experimental studies reported effects associated with the localization of the rotational axis in the molecular frame, including observations of nonergodic transitions in fullerene C$_{60}$~\cite{Liu_Science381_2023} and hyperfine-induced lifting of parity degeneracy in SF$_6$~\cite{Borde_PRL45_1980}, PH$_3$~\cite{Butcher_PRL70_1993}, and SiF$_4$~\cite{Pfister_PRL76_1996}.

Previous theoretical studies and analyses of molecules with localized rotations were primarily focused on characterizing their energy levels and spectroscopic transitions.
In this work, we extend these studies by investigating the effects of nuclear spin hyperfine interactions in rotational cluster states through first-principles calculations for the H$_2$S molecule.
Our calculations reveal that hyperfine interactions break nuclear spin symmetry, resulting in nearly equal mixing of \emph{ortho} and \emph{para} nuclear spin states of hydrogen atoms, with splittings on the order of MHz.
These represent the largest \emph{ortho}-\emph{para} splittings predicted for similar hydride molecules, for example, water~\cite{Yachmenev_JCP156_2022}.
The \emph{ortho}-\emph{para} coupling induces strong polarization of nuclear spins along the axis of rotation, with opposite orientations for hyperfine-split bands.
This behavior is similar to the Rashba effect, where the electron spin-orbit coupling arises due to inversion symmetry breaking~\cite{Bychkov_JETPLett39_1984, LaShell_PRL77_1996}.
Additionally, we predicted the hyperfine-induced lifting of degeneracy in states of opposite parity, a phenomenon previously observed in several molecules~\cite{Borde_PRL45_1980, Butcher_PRL70_1993, Pfister_PRL76_1996}.
This effect is explained using a simple model based on tunneling between rotating enantiomers, enabled by the Pauli exclusion principle.
The overall impact of hyperfine interactions in rotational cluster states can be understood using the concept of spin-induced diastereomers, providing a broader perspective on these effects in dynamically and statically chiral molecules.

We performed high-accuracy variational calculations of rovibrational states and hyperfine interactions in H$_2$S, employing the exact rovibrational kinetic energy operator and a high-accuracy spectroscopically refined potential energy surface (PES)~\cite{Azzam_MNRAS460_2016}.
Details of the vibrational coordinates, basis set, hyperfine interaction, and variational solution are provided in the supplementary information.
The electronic structure calculations of hyperfine spin-rotation tensor employed all-electron CCSD(T) method, as implemented in the quantum chemistry package CFOUR~\cite{cfour, Scuseria_JCP94_1991, Gauss_JCP105_1996, Gauss_MolPhys91_1997}, with the augmented core-valence correlation-consistent basis sets aug-cc-pwCVTZ~\cite{Peterson_JCP117_2002} for sulfur and aug-cc-pVTZ~\cite{Dunning_JCP90_1989, Kendall_JCP96_1992} for hydrogen atoms.
A complete documented Python implementation of all calculation steps is available in a public repository~\cite{github_rotational_clusters}.

\begin{figure}
\centering
\includegraphics[width=\linewidth]{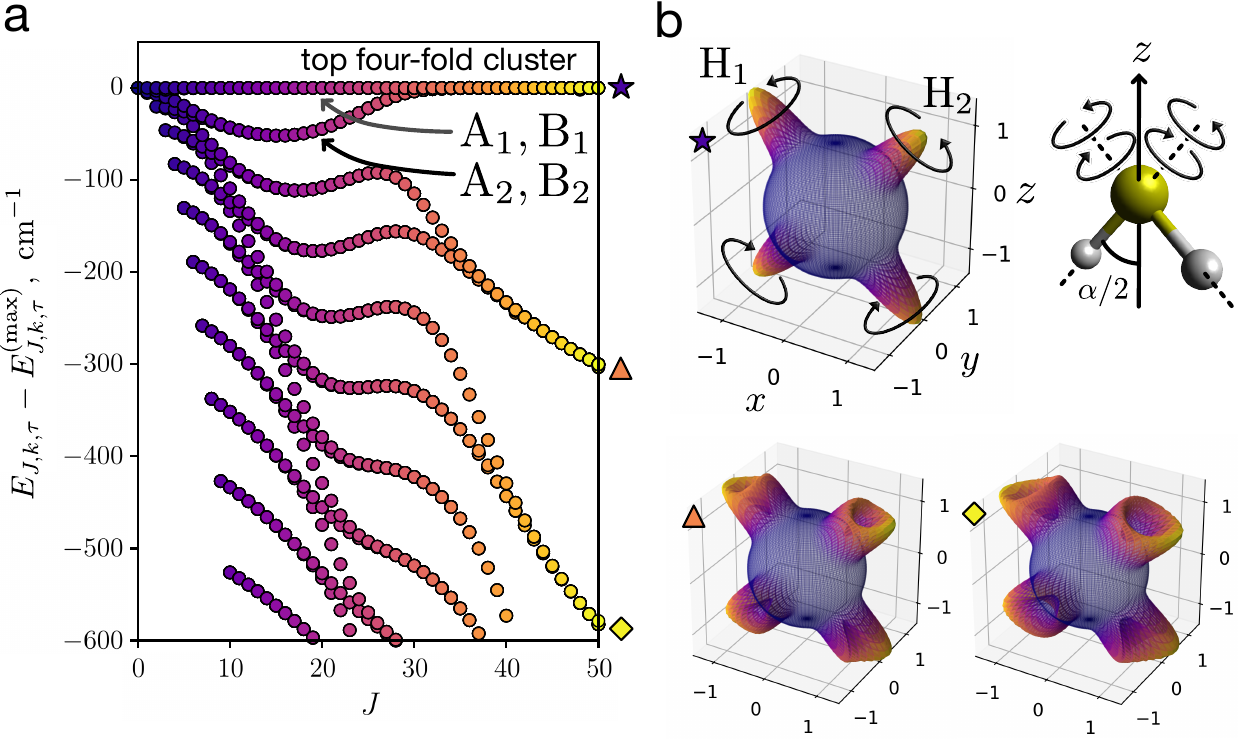}
\caption{
Rotational energy level clustering in the ground vibrational state of H$_2$S.
(a) Energy differences $E_{J,k,\tau}-E_{J,k,\tau}^{\text{(max)}}$ (\invcm) plotted for each rotational energy level $E_{J,k,\tau}$ relative to the maximum energy $E_{J,k,\tau}^{\text{(max)}}$ within its $J$ multiplet.
The rotational states group into clusters of four, corresponding to the A$_1$, A$_2$, B$_1$, and B$_2$ symmetries in \textbf{C}$_\text{2v}$.
(b) Reduced rotational probability density distributions of the rotational axis in the molecular frame for three groups of cluster states at $J=50$ and $|m|=J$.
The molecular-frame $z$-axis is defined to bisect the $\angle$H$_1$--S--H$_2$ valence angle.
The four maxima correspond to left- or right-handed rotation about axes closely aligned with the S--H$_1$ or S--H$_2$ bond.
}\label{fig1}
\end{figure}

An effective way to visualize rotational cluster states is by plotting rotational energies for different total angular momentum quantum numbers $J$, with energies shifted relative to the highest energy within a given multiplet, as shown in \autoref{fig1}.a.
A distinct characteristic of rotational cluster states is the localization of the rotational axis along the S--H bonds, which enables their clear identification.
This property is manifested in the reduced rotational probability density, as function of two Euler angles $\rho(\theta,\chi)=\int|\Psi|^2dq_1dq_2...dq_{3N-6}d\phi\sin\theta$, where the integration averages over all vibrational modes ($q_1,q_2,...,q_{3N-6}$) and the rotation ($\phi$) around the laboratory $Z$-axis.
In \autoref{fig1}.b, the normalized densities $\rho(\theta,\chi)$, plotted on a unit sphere, illustrate the distribution of the rotational axis in the molecular frame for three groups of cluster states.
The maxima in the plots correspond to localized rotations about axes closely aligned with the S--H bonds, with two oppositely located maxima along each bond corresponding to left- and right-handed rotations.

\begin{figure}
\centering
\includegraphics[width=\linewidth]{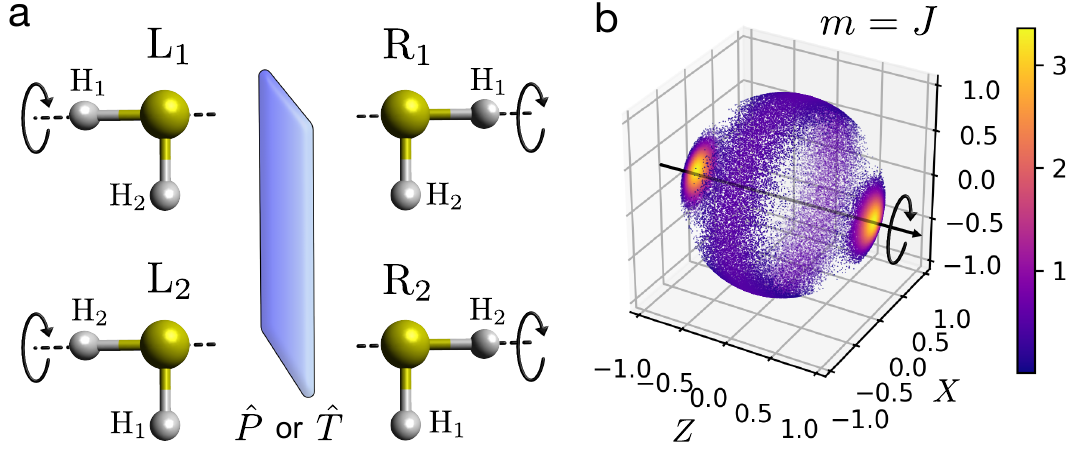}
\caption{
(a) Classical representation of rotating enantiomers in the cluster states of H$_2$S.
(b) Probability distribution of hydrogen nuclei in the laboratory frame, plotted in Cartesian coordinates on a unit sphere, for a selected cluster state of H$_2$S at $J=50$ and $m=J$.
}\label{fig2}
\end{figure}

The chirality of cluster states emerges as a result of breaking of parity (and time) reversal symmetry, combined with the permutational isolation of the two identical hydrogen nuclei.
Tunneling between enantiomers is suppressed by high kinetic energy barriers when the molecule is in a high angular momentum state.
The four cluster states of a triatomic \textbf{C}$_\text{2v}$-symmetry molecule, characterized by A$_1$, A$_2$, B$_1$, and B$_2$ symmetries, correspond to racemic mixtures of the four rotating enantiomers L$_1$, R$_1$, L$_2$, and R$_2$, representing left- (L) and right-handed (R) rotations about the S--H$_1$ or S--H$_2$ bond:
\begin{align}\label{eq:cluster_states}
|\text{A}_1\rangle=\left[(\text{R}_1+\text{R}_2)+(\text{L}_1+\text{L}_2)\right]/2, \\ \nonumber
|\text{A}_2\rangle=\left[(\text{R}_1+\text{R}_2)-(\text{L}_1+\text{L}_2)\right]/2, \\ \nonumber
|\text{B}_1\rangle=\left[(\text{R}_1-\text{R}_2)-(\text{L}_1-\text{L}_2)\right]/2, \\ \nonumber
|\text{B}_2\rangle=\left[(\text{R}_1-\text{R}_2)+(\text{L}_1-\text{L}_2)\right]/2.
\end{align}
By fixing the direction of rotation in the laboratory frame, the rotating enantiomers correspond to opposite orientations of one of the S--H bonds along the rotational axis, as illustrated in \autoref{fig2}.a.
The excitation of unidirectional rotation can be achieved using an optical centrifuge~\cite{Karczmarek_PRL82_1999, Korobenko_PRL116_2016}, which was also demonstrated for rotational cluster states~\cite{Owens_PRL121_2018,Owens_JPCL9_2018,Zak_PRR3_2021}.
\autoref{fig2}.b shows the probability distribution of hydrogen nuclei in the laboratory frame, plotted on a unit sphere for a cluster state at $J=50$ and $m=J$.
The distribution illustrates the spatial isolation of the two hydrogen nuclei: one is located at the poles of the sphere, while the other is distributed along a ring, representing a superposition of two classical left- and right-handed rotations about an axis closely aligned with one of the S--H bonds.
Tunneling between enantiomers associated with rotation about the same bond, i.e., $\text{L}_1\to\text{R}_1$ and $\text{L}_2\to\text{R}_2$, requires reversing the direction of rotation, which is energetically unfavorable.
However, since the two hydrogen atoms are identical, an alternative pathway $\text{L}_1\to\text{R}_2$ and $\text{L}_2\to \text{R}_1$ exist, which is more energetically feasible.
This transition can be achieved through a rotation of the molecule by about $90^\circ$ around an axis perpendicular to the main rotational axis.

An additional symmetry breaking effect arises in rotational cluster states due to differences in the internal magnetic field environments experienced by the two identical hydrogen nuclei, which rotate on distinctly different orbits.
The interaction of magnetic field, induced by molecular rotation, with the nuclear spins $I=1/2$ of the hydrogen nuclei (sulfur $^{32}$S has spin zero) is described by the spin-rotation coupling.
We neglect the spin dipole-dipole interaction, which is typically weaker and does not contribute to nuclear spin symmetry breaking in triatomic molecules~\cite{Yachmenev_JCP156_2022}.
The total spin-rovibrational wavefunctions $|F,m_F\rangle$ are built as symmetry-adapted combinations of the coupled products of rovibrational wavefunctions $|J,m_J\rangle$ and nuclear spin functions, \emph{para} $|I=0,m_I\rangle$ and \emph{ortho} $|I=1,m_I\rangle$.
Here, $J$, $I$, and $F$ are the quantum numbers of rotational, total nuclear spin, and total angular momentum operators, respectively, with $m_J$, $m_I$, and $m_F$ denoting their corresponding projections on the laboratory $Z$-axis.

\begin{figure}[b!]
\centering
\includegraphics[width=\linewidth]{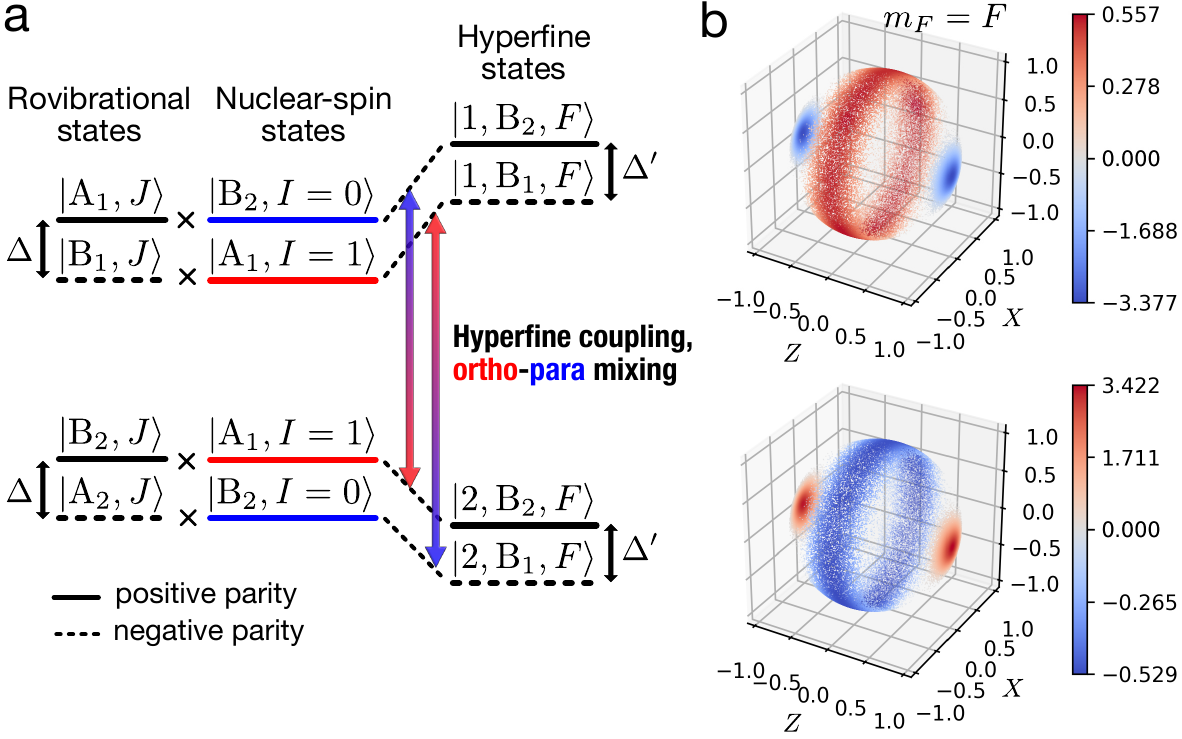}
\caption{
(a) Energy level diagram of H$_2$S, illustrating the coupling of rotational cluster states with \emph{para} ($I=0$) and \emph{ortho} ($I=1$) nuclear spin states of the hydrogen atoms.
Hyperfine interactions break spin symmetry, splitting states of the same parity and also lifting the original degeneracy ($\Delta\approx 0$) between opposite-parity states ($\Delta'>0$).
(b) This interaction polarizes the nuclear spins, resulting in opposite spin orientations along the rotational axis for states of the same parity, as illustrated by the computed spin-density distributions in the laboratory frame.
}\label{fig3}
\end{figure}

The symmetrization postulate requires the total wavefunction of the H$_2$S molecule to change sign upon exchange of the hydrogen nuclei, meaning it must transform as one of the irreducible representations B$_1$ or B$_2$ of the \textbf{C}$_\text{2v}$ symmetry group.
Accordingly, the \emph{ortho} spin state $|I=1\rangle$ of A$_1$ symmetry can form a symmetry-allowed product with the rovibrational states of B$_1$ and B$_2$ symmetries.
Similarly, the \emph{para} spin state $|I=0\rangle$ of B$_2$ symmetry can form a product with the rovibrational states of A$_1$ and A$_2$ symmetries.
The energy level diagram in \autoref{fig3}.a illustrates this coupling of rotational angular momentum with nuclear spin for the four cluster states.
In the absence of hyperfine interactions, the cluster states consist of two pairs of nearly exactly degenerate states of opposite parity, corresponding to rovibrational symmetries A$_1$ and B$_1$, as well as B$_2$ and A$_2$.
This near-degeneracy arises even at modest rotational excitations due to close proximity of the H$_2$S molecule to a near-oblate symmetric top.

The hyperfine interactions couple states of the same parity but with different total nuclear spin, i.e., $|\text{A}_1,J\rangle|I=0\rangle$ with $|\text{B}_2,J\rangle|I=1\rangle$ and $|\text{B}_1,J\rangle|I=1\rangle$ with $|\text{A}_2,J\rangle|I=0\rangle$.
As a result, states of the same parity split, breaking the nuclear spin symmetry with nearly equal mixing of \emph{ortho} and \emph{para} components.
The corresponding contributions to splittings increase from 514~kHz at $J=50$ to 2.2~MHz at $J=60$, whereas the base splittings between same-parity cluster states (i.e., without considering hyperfine interactions) decrease from 3.1~MHz at $J=50$ down to 1.5~KHz at $J=60$.
Analysis of spin-rotation coupling matrix elements reveals that cluster states exhibit significantly larger coupling elements compared to other states.
Interestingly, cluster states also appear in vibrationally excited states, with the spin-rotation coupling matrix elements of comparable magnitude to those in the ground-state cluster sates, resulting in similarly strong \emph{ortho}-\emph{para} mixing.
Rotational density plots for these states, along with a table of hyperfine coupling matrix elements, energies, splittings, and mixing coefficients, are provided in the supplementary information.

The breaking of nuclear spin symmetry induces spin polarization, with nuclear spin projections oriented along the rotational axis in the opposite directions for the two split energy bands. \autoref{fig3}.b illustrates this, showing calculated spin-densities of hydrogen nuclei in the laboratory frame, plotted in Cartesian coordinates on a unit sphere.
The spin-densities are calculated as $\boldsymbol{\rho}_l(\mathbf{r})=\sum_{i=1}^2\langle l,\text{B}_1,F|\delta(\mathbf{r}_{\text{H}_i}-\mathbf{r})\hat{\mathbf{I}}_{\text{H}_i}|l,\text{B}_1,F\rangle$, for two ($l=1,2$) hyperfine cluster states of B$_1$ symmetry and $m_F=F$, corresponding to unidirectional rotation of the molecule about the laboratory $Z$-axis.
The spin-density has only a non-zero projection along the rotational axis.
Notably, the spin-polarization vectors show opposite orientations for the hydrogen nucleus in the rotating bond and the hydrogen nucleus in the S--H bond about which the molecule rotates.

We analyze the effect of hyperfine coupling in terms of localized states representing rotating enantiomers.
The relationship between rotating enantiomers and symmetry-adapted delocalized cluster states is given in \eqref{eq:cluster_states}.
Each cluster state has a well-defined parity and is associated with either an \emph{ortho} or \emph{para} component of the total nuclear spin.
Hyperfine interactions preserve parity but couple states with different spin components, resulting in the following linear combinations of rotating enantiomers:
\begin{align}\label{eq:hyperfine_states}
|1,\text{B}_2\rangle = (\text{L}_1+\text{R}_1)|\uparrow\rangle|\downarrow\rangle+(\text{L}_2+\text{R}_2)|\downarrow\rangle|\uparrow\rangle,\\ \nonumber
|1,\text{B}_1\rangle = (\text{L}_1-\text{R}_1)|\uparrow\rangle|\downarrow\rangle+(\text{L}_2-\text{R}_2)|\downarrow\rangle|\uparrow\rangle, \\ \nonumber
|2,\text{B}_2\rangle = (\text{L}_1+\text{R}_1)|\downarrow\rangle|\uparrow\rangle+(\text{L}_2+\text{R}_2)|\uparrow\rangle|\downarrow\rangle,\\ \nonumber
|2,\text{B}_1\rangle = (\text{L}_1-\text{R}_1)|\downarrow\rangle|\uparrow\rangle+(\text{L}_2-\text{R}_2)|\uparrow\rangle|\downarrow\rangle.
\end{align}
Here, $|\uparrow\rangle|\downarrow\rangle$ and $|\downarrow\rangle|\uparrow\rangle$ represent the products of spin states of individual hydrogen nuclei, $|I_1,m_{I_1}\rangle|I_2,m_{I_2}\rangle = |\frac{1}{2},\frac{1}{2}\rangle|\frac{1}{2},-\frac{1}{2}\rangle$ and $|\frac{1}{2},-\frac{1}{2}\rangle|\frac{1}{2},\frac{1}{2}\rangle$, respectively. 
These states arise from symmetric and antisymmetric combinations of \emph{para} ($|I,m_I\rangle=|0,0\rangle$) and \emph{ortho} ($|I,m_I\rangle=|1,0\rangle$) total-spin states produced by hyperfine interactions (see diagram in \autoref{fig3}.a).
Specifically, the sum and difference of \emph{para} and \emph{ortho} states yield: $|0,0\rangle+|1,0\rangle=\sqrt{2}|\frac{1}{2},\frac{1}{2}\rangle|\frac{1}{2},-\frac{1}{2}\rangle$, $|0,0\rangle-|1,0\rangle=-\sqrt{2}|\frac{1}{2},-\frac{1}{2}\rangle|\frac{1}{2},\frac{1}{2}\rangle$.
The expressions in \eqref{eq:hyperfine_states} illustrate the spin polarization of the two hydrogen nuclei, with opposite nuclear spin projections along the axis of rotation.
States with the same parity, such as $|1,\text{B}_2\rangle$ and $|2,\text{B}_2\rangle$, or $|1,\text{B}_1\rangle$ and $|2,\text{B}_1\rangle$, also have different relative orientations of the nuclear spins, which leads to the energy splitting between them.
This difference is demonstrated by the calculated spin-densities for hyperfine-split bands, shown in \autoref{fig3}.b.

\begin{figure}
\centering
\includegraphics[width=\linewidth]{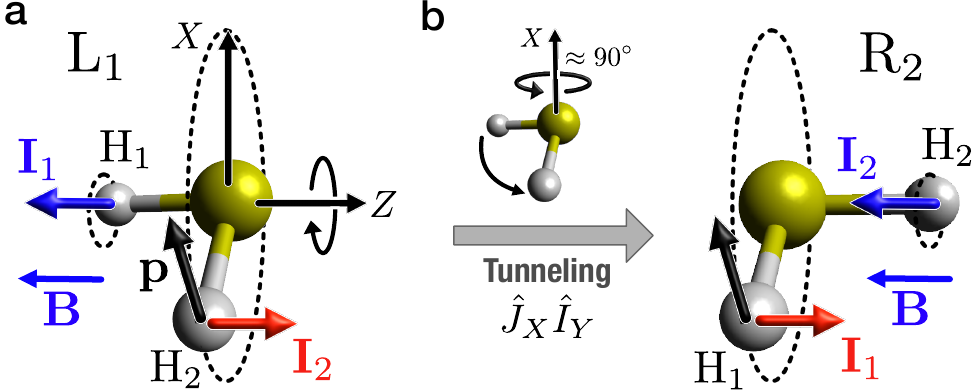}
\caption{
(a) A schematic, classical depiction of the internally-induced magnetic field generated by the localized rotation of H$_2$ about S--H$_1$ bond in $L_1$ rotating enantiomer of H$_2$S.
The magnetic field polarizes the nuclear spin of H$_1$.
Due to non-overlapping spatial distributions of the H$_1$ and H$_2$ nuclei, the Pauli exclusion principle requires the spin of H$_2$ to be polarized in the opposite direction of H$_1$ to maintain overall antisymmetry.
(b) The energetically most feasible tunneling pathway between two rotating enantiomers involves a rotation of approximately $90^\circ$ around an axis perpendicular to the main rotation axis, converting $L_1$ into $R_2$ and exchanging the spins of H$_1$ and H$_2$.
}\label{fig4}
\end{figure}

Nuclear spin polarization can be understood using classical arguments.
Consider the hydrogen atom H$_2$ rotating around an axis aligned with the S--H$_1$ bond.
This rotation generates a magnetic field at the position of the H$_1$ nucleus on the rotational axis, polarizing its nuclear spin either parallel or antiparallel to the field direction, as illustrated in \autoref{fig4}.a.
The two hydrogen nuclei, one revolving on a larger orbit (H$_2$) and the other located on the rotational axis (H$_1$), are permutationally isolated.
Hence, the spatial part of the total wavefunction is symmetric with respect to the exchange of the two hydrogen nuclei.
To satisfy the Pauli exclusion principle, the antisymmetric character of the total wavefunction must reside in the spin component.
As a result, the spin of the rotating nucleus, H$_2$, must be polarized in the opposite direction to that of the H$_1$ nucleus on the rotational axis.

As shown in \autoref{fig3}.a, the hyperfine interaction repels (couples) states of the same parity, each of which has a degenerate state component of opposite parity.
The coupling results in a hyperfine energy splitting that is slightly different for the positive-parity state pairs as compared to the negative-parity state pairs, effectively lifting the original degeneracy between opposite-parity states.
For example, at $J=50$, positive-parity B$_2$ symmetry cluster states of H$_2$S split by 3.613~MHz, while the negative-parity B$_1$ symmetry states split by 3.640~MHz.
This 27~kHz difference causes two pairs of B$_1$ and B$_2$ cluster states, originally degenerate within 0.2~kHz, to shift apart by about 13.5~kHz.
A similar effect was observed in high-precision spectroscopic measurements of the $\nu_4$ vibrational band of PH$_3$~\cite{Butcher_PRL70_1993} and the $\nu_3$ band of SiH$_4$~\cite{Pfister_PRL76_1996} and SF$_6$~\cite{Borde_PRL45_1980}, where doubly degenerate E-symmetry states were split by 8--32~kHz.
The authors attributed this lifting of degeneracy to hyperfine mixing with nearby states, which introduced slightly different interaction strengths for positive- versus negative-parity states.

This effect can be understood using the picture of tunneling between different rotating enantiomers, induced by hyperfine interactions.
From \eqref{eq:hyperfine_states} it follows that the energy differences between opposite-parity hyperfine cluster states, e.g., $|1,\text{B}_2\rangle$ and $|1,\text{B}_1\rangle$, arise from nonzero coupling between $\text{L}_1$ and $\text{R}_1$ (or $\text{L}_2$ and $\text{R}_2$), or between $\text{L}_1$ and $\text{R}_2$ (and similarly $\text{L}_2$ and $\text{R}_1$).
As noted earlier, the $\text{L}_1\to\text{R}_2$ transition, which involves a rotation of approximately $90^\circ$ around an axis perpendicular to the main rotational axis, is energetically more favorable than the $\text{L}_1\to\text{R}_1$ transition.
This motion effectively exchanges H$_1$ and H$_2$ combined with an inversion operation.
Since the direction of rotation remains unchanged, the magnetic field induced at the nucleus located on the rotational axis  remains the same, aligning the spin in the same direction.
The spin of the rotating nucleus also remains oppositely aligned, as shown in \autoref{fig4}.b.
The H$_1$ and H$_2$ however alternate between being on the rotational axis and on an orbiting trajectory, which swaps their nuclear spin projections.
Such transition can be facilitated by the spin-rotation coupling element, such as $\hat{J}_X \hat{I}_Y$.
Here, the $\hat{I}_X$ operator flips the spins, while the $\hat{J}_Y$ alters the projection of the rotational angular momentum, resulting in an enhanced precession of the S--H bond about the rotational axis, which favours the $\text{L}_1\to\text{R}_2$ transition.

Interestingly, if the Pauli exclusion principle were disregarded, the spin polarization of the proton on an orbiting trajectory would be zero, and the tunneling $\text{L}_1\to\text{R}_2$ depicted \autoref{fig4}.b would be forbidden.
In this case, the degeneracy of opposite-parity states would not be lifted.
This conclusion aligns with experimental observations of hyperfine-induced parity degeneracy lifting~\cite{Borde_ChemPhys71_1982,Butcher_PRL70_1993,Pfister_PRL76_1996}, where the authors emphasize that this phenomenon can be seen as a consequence of the Pauli exclusion principle forbidding two degenerate states differing only by parity.

An external electric field can be used to couple cluster states of different parity and same spin polarization through dipole interaction, i.e., $|1,\text{B}_2\rangle$ with $|1,\text{B}_1\rangle$ and $|2,\text{B}_2\rangle$ with $|2,\text{B}_1\rangle$ (see \autoref{fig3}), to produce pure rotating enantiomers~\cite{Owens_PRL121_2018}.
In the absence of hyperfine interactions, the rotating enantiomers originating from rotations about the S--H$_1$ or S--H$_2$ bonds are indistinguishable.
However, hyperfine interactions introduce an additional chiral center - the nuclear spin, which, combined with rotational chirality, gives rise to two chiral diastereomers.
Much like chemical diastereomers, these states have different energies and also opposite orientations of the nuclear spin projections onto the rotational axis.
These diastereomers, however, cannot be assigned as rotating about a specific bond (i.e., $L_1$ or $L_2$), rather, they represent a coherent mixture of rotations about both S--H bonds.
The Stark energy plots for hyperfine cluster states, together with rotational spin-density distributions for the resulting rotating diastereomers, are provided in the supplementary information.

This work demonstrates nuclear spin symmetry breaking and nuclear spin polarization in the laboratory frame for the H$_2$S molecule, arising in rotational energy level clusters as a result of dynamic symmetry breaking.
This effect is not merely a result of near-degeneracy of cluster states, i.e., it does not occur in other similarly closely degenerate states with symmetry-allowed hyperfine interactions.
Instead, it emerges from the unique rotational character of these states, isolating identical nuclei on different, non-overlapping trajectories.
Hyperfine interactions enable tunneling between rotating enantiomers, a phenomenon intrinsically tied to the Pauli exclusion principle, explaining previous spectroscopic observations of parity-degeneracy lifting.
This effect may be a more general characteristic of hyperfine interactions in chiral molecules.

Although populating rotational cluster states in small molecules like H$_2$S requires tailored laser fields~\cite{Owens_JPCL9_2018}, larger molecules, such as fullerene C$_{60}$, can significantly populate states with rotational axis localization and symmetry breaking even at room temperature~\cite{Liu_Science381_2023}.
We also predicted the existence of rotational cluster states at much lower angular momenta in vibrationally excited bending bands of H$_2$S, where bending vibrations facilitate the molecular distortion necessary for the formation of rotational clusters.

\section*{Acknowledgments}
We would like to thank our colleague Emil Vogt for insightful discussions and critical suggestions to this manuscript.
This work was supported by Deutsches Elektronen-Synchtrotron DESY, a member of the Helmholtz Association
(HGF), including the Maxwell computational resource operated at DESY.

\bibliography{hyperfine, rotational_clusters, symm_break_kinetic, chirality_true_false, chiral_field, CISS, Rashba_effect, PES, quantum_chemistry, codes, optical_centrifuge, theory_rovibrations, DMS}%
\onecolumngrid%
\listofnotes%

\section{Supplementary information}
\section{Theoretical details}

The spin-rotation coupling is the interaction between the rotational angular momentum $\mathbf{J}$ of the molecule and the nuclear spins $\mathbf{I}_n$ of different nuclei~\cite{Flygare_ChemRev74_1974}
\begin{align}\label{eq:sr_cart}
\hat{H}_\text{sr} = \sum_{n}^{N_I} \mathbf{I}_n \cdot \mathbf{M}_n \cdot \mathbf{J},
\end{align}
where $\mathbf{M}_n$ is the second-rank spin-rotation tensor relative to the nucleus $n$ and the sum runs over a number of nuclei $N_I$ with nonzero spin.
Using the spherical-tensor representation~\cite{Zare_AngularMomentum}, the spin-rotation Hamiltonian can be expressed as
\begin{align}\label{eq:sr_spher}
\hat{H}_\text{sr} = &\frac{1}{2}\sum_{n}^{N_I} \sum_{\omega=0}^2 \sqrt{2\omega+1} \left(-\frac{1}{\sqrt{3}}\right) \mathbf{I}_n^{(1)}
\cdot\left( (-1)^\omega \left[ \mathbf{M}_n^{(\omega)} \otimes \mathbf{J}^{(1)} \right]^{(1)} + \left[ \mathbf{J}^{(1)} \otimes \mathbf{M}_n^{(\omega)} \right]^{(1)} \right),
\end{align}
where $\mathbf{M}_n^{(\omega)}$, $\mathbf{J}^{(1)}$, and $\mathbf{I}_n^{(1)}$ denote the spherical-tensor representations of operators in \eqref{eq:sr_cart} and the square brackets are used to indicate the tensor product of two spherical-tensor operators.
Because the spin-rotation tensor is not necessarily symmetric, the second term in the sum \eqref{eq:sr_spher} is included to ensure that the Hamiltonian remains Hermitian.

The nuclear-spin operators $\mathbf{I}_n$ of the two hydrogen nuclei in H$_2$S and the rotational angular momentum operator $\mathbf{J}$ are coupled as $\mathbf{I}=\mathbf{I}_1+\mathbf{I}_2$ and $\mathbf{F}=\mathbf{J}+\mathbf{I}$.
The nuclear-spin functions $\ket{I,m_I}$ depend on the quantum numbers $I$ and $m_I$ of the collective nuclear-spin operator $\mathbf{I}$ and its projection onto the laboratory $Z$-axis, respectively.
The total spin-rovibrational wavefunctions $\ket{F,m_F,u}$ are built as symmetry-adapted linear combinations of the coupled products of the rovibrational wavefunctions $\ket{J,m_J,l}$ and the nuclear-spin functions $\ket{I,m_I}$.
Here, $J$ and $F$ are the quantum numbers of $\mathbf{J}$ and $\mathbf{F}$ operators with $m_J$ and $m_F$ of their laboratory $Z$-axis projections.
Here, $l$ and $u$ represent the rovibrational and hyperfine state indices, respectively, encompassing all relevant quantum numbers required to fully characterize a spin-rovibrational state.
These include rotational $k$ and vibrational quantum numbers $v_1,v_2,v_3$, symmetries of the total wavefunction and its rotational, vibrational, and nuclear-spin components.

The matrix representation of the spin-rotation Hamiltonian in the basis of the $\ket{F,m_F,u}$ functions is diagonal in $F$ and $m_F$, with the explicit expressions given by
\begin{align}\label{eq:sr_me}
\langle &F,m_F,u'|\hat{H}_\text{sr}|F,m_F,u\rangle
=\frac{1}{2}(-1)^{I+F}\sqrt{(2J+1)(2J'+1)} \left\{ \begin{array}{ccc}I'&J'&F\\J&I&1\end{array}\right\} \\ \nonumber
&\times\sum_{n}^{N_I}\sum_{\omega=0}^2 N_\omega
\left[
(-1)^\omega J \left\{ \begin{array}{ccc}\omega&1&1\\J&J'&J\end{array}
\right\}\left( \begin{array}{ccc}J&1&J\\-J&0&J\end{array} \right)^{-1}
+J' \left\{ \begin{array}{ccc}1&\omega&1\\J&J'&J'\end{array}\right\}\left(
\begin{array}{ccc}J'&1&J'\\-J'&0&J'\end{array} \right)^{-1}\right]
\mathcal{M}_{\omega,n}^{(J',l',J,l)} \langle I'||\mathbf{I}_n^{(1)}||I\rangle,
\end{align}
where the normalization constant $N_\omega=1$, $-\sqrt{3}$, and $\sqrt{5}$ for $\omega=0$, 1, and 2, respectively.
The expressions for the reduced matrix elements of the nuclear-spin operators
$\langle I'||\mathbf{I}_n^{(1)}||I\rangle$ depend on the total number of coupled spins and can be computed using a general recursive procedure as described, for example, in~\cite{Yachmenev_JCP147_2017}.
Here, for the two equivalent hydrogen spins $I_1=I_2=1/2$, the reduced matrix elements are
\begin{align}\label{eq:spin_red_me}
& \langle I'||\mathbf{I}_n^{(1)}||I\rangle = (-1)^{I\delta_{n,1}+I'\delta_{n,2}} I_1 
\sqrt{(2I+1)(2I'+1)}\left\{ \begin{array}{ccc}I_1&I'&I_1\\I&I_1&1\end{array} \right\}
\left( \begin{array}{ccc}I_1&1&I_1\\-I_1&0&I_1\end{array} \right)^{-1},
\end{align}
with the explicit values $\langle 0||\mathbf{I}_n^{(1)}||0\rangle=0$, $\langle 1||\mathbf{I}_n^{(1)}||1\rangle=\sqrt{3/2}$, $\langle 0||\mathbf{I}_n^{(1)}||1\rangle=\pm\sqrt{3}/2$ for $n=1$ and $2$, respectively, and $\langle 1||\mathbf{I}_n^{(1)}||0\rangle=\mp\sqrt{3}/2$.

The expression for the rovibrational matrix elements of spin-rotation tensor $\mathcal{M}_{\omega,n}^{(J',l',J,l)}$ in \eqref{eq:sr_me} depends on the rovibrational wavefunctions $\ket{J,m_J,l}$, which are represented by linear combinations of products of vibrational wavefunctions $\ket{\nu}=\ket{v_1,v_2,v_3}$ and symmetric-top rotational functions
\begin{align}\label{eq:rovib_wf}
|J,m_J,l\rangle = \sum_{\nu,k} c_{\nu,k}^{(J,l)}\ket{\nu}\ket{J,k,m_J}.
\end{align}
In this basis, the matrix elements $\mathcal{M}_{\omega,n}^{(J',l',J,l)}$ can be expressed as
\begin{align}\label{eq:m_tens}
& \mathcal{M}_{\omega,n}^{(J',l',J,l)} = \sum_{\nu',k'}\sum_{\nu,k}
c_{\nu',k'}^{(J',l')*}\, c_{\nu,k}^{(J,l)} \, (-1)^{k'}
\sum_{\sigma=-\omega}^\omega\sum_{\alpha,\beta=x,y,z}
\left(\begin{array}{ccc} J & \omega & J' \\ k & \sigma & -k'\end{array}\right)
U_{\omega\sigma,\alpha\beta}^{(2)} \langle \nu'|\bar{M}_{\alpha\beta,n}|\nu\rangle,
\end{align}
where $\bar{M}_{\alpha\beta,n}$ is spin-rotation tensor in the molecule-fixed frame and the $9\times9$ constant matrix $U_{\omega\sigma,\alpha\beta}^{(2)}$ ($\omega=0,\dots,2$, $\sigma=-\omega,\ldots,\omega$) defines the transformation of a general second-rank Cartesian tensor operator into its spherical-tensor representation, see, e.g., (5.41)--(5.44) in~\cite{Zare_AngularMomentum}.

The total Hamiltonian $\hat{H}$ is composed of the pure rovibrational Hamiltonian $\hat{H}_\text{rv}$ and the hyperfine term $\hat{H}_\text{sr}$.
In the basis of wavefunctions defined in \eqref{eq:rovib_wf}, the rovibrational Hamiltonian $\hat{H}_\text{rv}$ is diagonal, with its elements corresponding to rovibrational energies.
The matrix elements of the total Hamiltonian are given by:
\begin{align}\label{eq:tot_ham}
\langle F,m_F,u'|\hat{H}|F,m_F,u\rangle &= E_u\delta_{J,J'}\delta_{l,l'}\delta_{I,I'}
+ \langle F,m_F,u'|\hat{H}_\text{sr}|F,m_F,u\rangle.
\end{align}

\section{Computational details}

The spin-rotation tensors  in the molecule-fixed frame, $\bar{\mathcal{M}}_{\alpha\beta,n}$ (in \eqref{eq:m_tens}), were calculated at the equilibrium geometry of the molecule ($r_{\text{SH}} = 1.35$~\AA\ and $\alpha_{\angle\text{HSH}}= 92^\circ$) using all-electron CCSD(T) method, as implemented in the quantum chemistry package CFOUR~\cite{cfour, Scuseria_JCP94_1991, Gauss_JCP105_1996, Gauss_MolPhys91_1997}, with the augmented core-valence correlation-consistent basis sets aug-cc-pwCVTZ~\cite{Peterson_JCP117_2002} for sulfur and aug-cc-pVTZ~\cite{Dunning_JCP90_1989, Kendall_JCP96_1992} for hydrogen atoms.
\autoref{tab:cfour_sr} presents elements of the spin-rotation tensors, with the corresponding atomic Cartesian coordinates provided in~\autoref{tab:atom_xyz}.
In the variational calculations, the spin-rotation tensors were transformed into the molecule-fixed frame used for the construction of the kinetic energy operator.

\begin{table}
\centering
\renewcommand{\arraystretch}{1.2}
\begin{tabular*}{0.5\linewidth}{@{\extracolsep{\fill}}lrrr}
\hline
& $J_x$ & $J_y$ & $J_z$ \\
\hline
$x$(H$_1$) &  16.747 &   0 &   0 \\
$y$(H$_1$) &   0 &  17.423 & -18.341 \\
$z$(H$_1$) &   0 & -20.256 &  13.384 \\
\hline
$x$(H$_2$) &  16.747 &   0 &   0 \\
$y$(H$_2$) &   0 &  17.423 &  18.341 \\
$z$(H$_2$) &   0 &  20.256 &  13.384 \\
\hline
\end{tabular*}
\caption{Spin-rotation tensors $\bar{\mathcal{M}}_{\alpha\beta,n}$ (kHz) of H$_2$S in the molecule-fixed frame, calculated at equilibrium geometry $r_{\text{SH}} = 1.35$~\AA\ and $\alpha_{\angle\text{HSH}}= 92^\circ$. The corresponding Cartesian coordinates of atoms are listed in \autoref{tab:atom_xyz}}\label{tab:cfour_sr}
\end{table}

\begin{table}
\centering
\renewcommand{\arraystretch}{1.2}
\begin{tabular*}{0.5\linewidth}{@{\extracolsep{\fill}}lrrr}
\hline
S    &           0.00000000  &   0.00000000 &    0.10509864 \\
H    &           0.00000000  &  -1.83512992 &   -1.66706573 \\
H    &           0.00000000  &   1.83512992 &   -1.66706573 \\
\hline
\end{tabular*}
\caption{Reference equilibrium Cartesian coordinates of atoms (in Bohr)}\label{tab:atom_xyz}
\end{table}

To represent vibrational motions of H$_2$S, valence internal coordinates were used.
The molecular frame was defined such that the $z$-axis bisects the valence angle and the molecule lies in the $xz$ plane.
The rovibrational kinetic energy operator was calculated exactly using the procedure outlined in~\cite{Matyus_JCP130_2009, Yachmenev_JCP143_2015}.
For the potential energy surface (PES), the spectroscopically refined PES of H$_2$S was employed~\cite{Azzam_MNRAS460_2016}.
For simulations of the Stark effect, the \emph{ab initio} dipole moment surface of H$_2$S from~\cite{Azzam_JQSRT161_2015} was used.

Hermite functions $\mathcal{H}_i(x)$ were utilized as the basis for all vibrational coordinates, with vibrational matrix elements computed using a direct product of 80th-order Gauss-Hermite quadratures for each mode.
To map the valence stretching and bending coordinates $\xi_1,\xi_2,\xi_3=r_{\text{SH}_1},r_{\text{SH}_2},\alpha_{\angle\text{H}_1\text{SH}_2}$  into the $(-\infty, \infty)$ range of the Hermite function argument  $x$, a linear mapping $\xi_l = a_lx_l + b_l$ ($l=1..3$) was applied, with
\begin{align}
a_l = \left(\frac{G_{l,l}}{F_l}\right)^{1/4},~b_l=\xi_\text{eq},
\end{align}
where the force constants $F_l=\partial^2V/\partial \xi_l^2$ and diagonal elements of the kinetic energy matrix $G_{l,l}$ both evaluated at the equilibrium values of the valence coordinates $\xi_\text{eq}$.

The rovibrational wavefunctions and energies were computed in the following steps.
In the first step, a contracted basis set $\psi^{(l)}$ was constructed by solving a set of eigenvalue problems independently for each vibrational mode $\xi_l$.
These eigenvalue problems are based on the Hamiltonian obtained by averaging the total vibrational Hamiltonian $\hat{H}_\text{vib}$ over all other vibrational modes, using a basis consisting of a product of zero-order Hermite functions.
The resulting contracted functions are expressed as:
\begin{align}
\psi_j^{(l)} = \sum_i c_{ij}^{(l)} \mathcal{H}_i(\xi_l),
\end{align}
where the expansion coefficients $c_{ij}^{(l)}$ are obtained by solving the eigenvalue problem:
\begin{align}
\mathbf{H}^{(l)}\mathbf{c}^{(l)} = \boldsymbol{\epsilon}^{(l)}\mathbf{c}^{(l)},
\end{align}
with matrix elements given by:
\begin{align}
H_{i',i}^{(l)}&=\left\langle \mathcal{H}_{i'}(\xi_l)\left|\hat{H}^{(l)}\right|\mathcal{H}_i(\xi_l)\right\rangle, \\
\hat{H}^{(l)}&=\left\langle \prod_{l'\neq l}\mathcal{H}_0(\xi_{l'})\left|\hat{H}_\text{vib}\right|\prod_{l'\neq l}\mathcal{H}_0(\xi_{l'})\right\rangle.
\end{align}

The total vibrational wavefunctions were expressed as linear combinations of direct products of contracted functions for each vibrational mode, truncated using a polyad condition, as
\begin{align}\label{eq:vib_wf}
|\nu\rangle =\sum_{2i+2j+k\leq P_\text{max}} c_{ijk,\nu}\psi_i^{(1)}(\xi_1)\psi_j^{(2)}(\xi_2)\psi_k^{(3)}(\xi_3).
\end{align}
The expansion coefficients $c_{ijk,\nu}$ and vibrational energies were determined by solving an eigenvalue problem for the total vibrational Hamiltonian $\hat{H}_\text{vib}$.
The size of the vibrational basis set was controlled by the maximum plyad number $P_\text{max}$.
In the present calculations, $P_\text{max}=20$ and 24 were used, ensuring convergence of hyperfine calculations to an accuracy below 1~kHz.

The rovibrational wavefunctions were expressed as linear combinations of the total vibrational wavefunctions from~\eqref{eq:vib_wf} and symmetric-top functions in the Wang representation, as defined in~\eqref{eq:rovib_wf}.
The rovibrational wavefunctions and energies were determined by solving an eigenvalue problem for the total rovibrational Hamiltonian $\hat{H}_\text{rv}$.
This was performed independently for different values of the rotational angular momentum quantum number $J$ and for different symmetry representations of the \textbf{C}$_\text{2v}$ symmetry group.

The computation of hyperfine energies and wavefunctions involved the following steps.
First, the rovibrational matrix elements of the spin-rotation tensors were calculated using the expression in~\eqref{eq:m_tens}.
These matrix elements were subsequently used to construct the spin-rotation Hamiltonian using~\eqref{eq:sr_me}.
The total Hamiltonian was then assembled as the sum of purely rovibrational part, which is diagonal and determined by the rovibrational energies, and a non-diagonal spin-rotation part, as expressed in~\eqref{eq:tot_ham}.
Finally, the hyperfine energies and wavefunctions were obtained by diagonalizing the total Hamiltonian.

In this work, we computed the rotational spin-density distributions to characterize spin polarization in hyperfine states.
The nuclear spin-density is given by:
\begin{align}
  \boldsymbol{\rho}({\mathbf{r}}) = \sum_{n=1}^{N_I} \delta(\mathbf{r_n}-\mathbf{r})\hat{\mathbf{I}}_n,
\end{align}
where $n=1..N_I$ denotes the nuclei with corresponding spin operators $\hat{\mathbf{I}}_n$, and $\mathbf{r}$ represents a set of rotational Euler angles.
The matrix elements of the spin-density between hyperfine wavefunctions can be expressed as:
\begin{align}
  \langle& F',m_F',u'|\boldsymbol{\rho}|F,m_F,u\rangle  = 
  \sum_{I',J',l'}\sum_{I,J,l} c_{I',J',l'}^{(F',u')*} \, c_{I,J,l}^{(F,u)}
  \sum_{m_I',m_I} G_{m_F',m_I',l'}^{(F',I',J')*}\, G_{m_F,m_I,l}^{(F,I,J)}\delta_{\nu'\nu} \langle I',m_I'|\hat{\mathbf{I}}^{(n)}|I,m_I\rangle,
\end{align}
where
\begin{align}
  G_{m_F,m_I,l}^{(F,I,J)}(\mathbf{r})&= (-1)^{F+m_F}\sqrt{2F+1}
  \sum_{m_J}\left(\begin{array}{ccc}F&I&J\\-m_F&m_I&m_J\end{array}\right)|J,m_J,l\rangle(\mathbf{r}),
\end{align}
and $c_{I,J,l}^{(F,u)}$ are hyperfine wavefunction coefficients obtained by diagonalization of the total Hamiltonian.
Here, we assume that the vibrational basis $|\nu\rangle$ from \eqref{eq:vib_wf} is orthonormal and integrate out the vibrational coordinates.
The matrix elements of the nuclear spin operators can be calculated using Wigner-Eckart theorem, with reduced matrix elements provided in \eqref{eq:spin_red_me}.
A Monte-Carlo procedure was employed to sample the spin-density over the rotational angles.

To simulate the Stark effect for hyperfine energy levels, it is necessary to compute the matrix elements of the permanent electric dipole moment operator in the laboratory frame.
This calculation involved the following steps.
First, the rovibrational matrix elements of the dipole moment surface are computed and expressed in a tensor form similar to \eqref{eq:m_tens}:
\begin{align}\label{k_tens}
\mathcal{K}^{(J',l',J,l)} = \sum_{\nu' k'}\sum_{\nu k}
c_{\nu',k'}^{(J',l')*} \, c_{\nu,k}^{(J,l)} \, (-1)^{k'}
\sum_{\sigma=-1}^1\sum_{\alpha,\beta=x,y,z}
\left(\begin{array}{ccc} J & 1 & J' \\ k & \sigma & -k'\end{array}\right)
U_{\sigma,\alpha}^{(1)} \langle \nu'|\bar{\mu}_{\alpha}|\nu\rangle,
\end{align}
where $\bar{\mu}_{\alpha}$ ($\alpha=x,y,z$) is the permanent dipole moment in the molecule-fixed frame and the $3\times3$ constant matrix $U_{\sigma,\alpha}^{(1)}$ ($\omega=1$, $\sigma=-\omega,\ldots,\omega$) defines the transformation of a rank-1 Cartesian tensor operator into its spherical-tensor representation, see, e.g., (5.4) in~\cite{Zare_AngularMomentum}.
In the second step, the dipole matrix elements are transformed into the basis of hyperfine wavefunctions:
\begin{align}
  \mathcal{K}^{(F',u',F,u)} = \sum_{I',J',l'}\sum_{I,J,l} c_{I',J',l'}^{(F',u')*} \, c_{I,J,l}^{(F,u)}
(-1)^{I+F+J+J'}\sqrt{(2J'+1)(2J+1)}
\left\{\begin{array}{ccc}J'&F'&I \\ F&J&1\end{array}\right\} \, \mathcal{K}^{(J',l',J,l)}\delta_{I',I}
\end{align}
where $c_{I,J,l}^{(F,u)}$ are hyperfine wavefunction coefficients obtained by diagonalization of the total Hamiltonian.
Finally, the matrix elements of the laboratory-frame dipole moment operator $\mu_A$ ($A=X,Y,Z$) are computed as:
\begin{align}
  \langle& F',m_F',u'|\mu_A|F,m_F,u\rangle  =
  (-1)^{F'-m_F'}\sqrt{(2F'+1)(2F+1)} \, \mathcal{K}^{(F',u',F,u)}
  \sum_{\sigma=-1}^{1}[U^{(1)}]^{-1}_{A,\sigma}
  \left(\begin{array}{ccc}F' & 1 & F \\ -m_F'&\sigma&m_F\end{array}\right).
\end{align}

\section{\emph{Ortho}-\emph{para} mixing in hyperfine cluster states}

To identify states and characteristic molecular rotations that lead to the strongest \emph{ortho}-\emph{para} mixing of spin states, it is logical to analyze the rovibrational states with the largest values of spin-rotation matrix element $\mathcal{M}_{\omega,n}^{(J',l',J,l)}$ in~\eqref{eq:m_tens}.
\autoref{tab:sr} presents such states with largest absolute values of ${M}_{\omega=2,n}^{(J',l',J,l)}$ matrix element for selected $J$ values.
The table also includes the corresponding plots of reduced rotational probability density functions for each selected state pair, calculated as:
\begin{align}
\rho(\theta,\chi)=\int\left||J,m_J,l\rangle\right|^2\sin\theta dq_1dq_2...dq_{3N-6}d\phi,
\end{align}
with $m_J=J$.
Here, the integration averages over all vibrational modes ($q_1,q_2,...,q_{3N-6}$) and the rotation ($\phi$) around the laboratory $Z$-axis.
The resulting rotational density distributions represent the probability of orientation of the rotational axis in the molecule-fixed frame.
This frame is defined by the $z$-axis bisecting the $\angle$H$_1$SH$_2$ valence angle, with the molecule placed in the $xz$ plane.

\autoref{tab:sr} lists the largest spin-rotation matrix elements between A$_1$ and B$_2$ state symmetry pairs.
For each $J$ value, the lowest-energy state corresponds to the ground vibrational state, while the higher-energy states correspond to the first and second excited vibrational bending bands, $\nu_2$ and $2\nu_2$.
For all states with large spin-rotation coupling matrix element, the rotational density distributions appear nearly identical, indicating localization of the rotational axis along directions aligned with the S--H bonds.
Similar results are obtained for the A$_2$ and B$_1$ state symmetry pairs.

These findings clearly demonstrate that strong spin-rotation interaction and spin symmetry breaking is a distinctive feature of the rotational cluster states, which exist even in vibrationally excited bands.
It is important to note that in this work, the electronic structure calculations of the spin-rotation tensors were performed only at the equilibrium geometry of the molecule, making them independent on vibrational motions.
While vibrational contributions to spin-rotation tensors are expected to be minimal for cluster states in the ground vibrational state, in vibrationally excited sates, incorporating vibrational dependence into the spin-rotation tensors is likely to enhance the corresponding rovibrational matrix elements $\mathcal{M}_{\omega,n}^{(J',l',J,l)}$.
This, in turn, would lead to a stronger \emph{ortho}-\emph{para} mixing even at relatively low rotational excitations.

\LTcapwidth=\textwidth
\renewcommand{\arraystretch}{1.2}
\begin{longtable*}{@{\extracolsep{\fill}}cccccccc}
\caption{Rovibrational states and corresponding reduced rotational density distributions with the largest spin-rotation matrix elements $\mathcal{M}_{2,1}=-\mathcal{M}_{2,2}$.}\label{tab:sr}
	\endfirsthead
	\endhead
	\hline
  $J=J'$ & $\Gamma'$ & $E'$, cm$^{-1}$ & $\Gamma$ & $E$, cm$^{-1}$& $\mathcal{M}_{2,1}$, kHz & $\rho'$ & $\rho$ \\
	\hline
  40 & A$_1$ & 20344.2209598 & B$_2$ & 20344.2206555&  -2.4 & \adjustbox{valign=m}{\includegraphics[width=2cm]{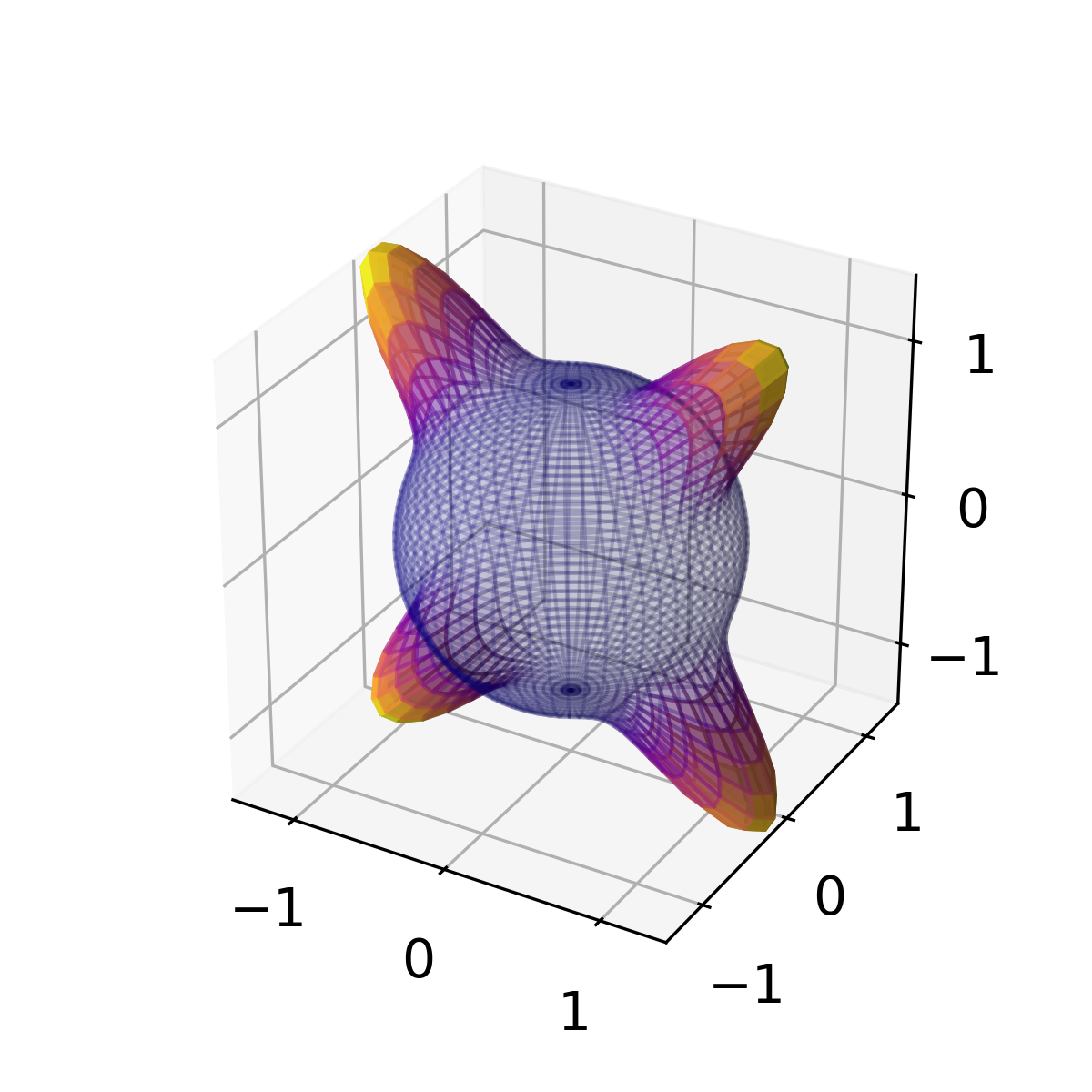}} & \adjustbox{valign=m}{\includegraphics[width=2cm]{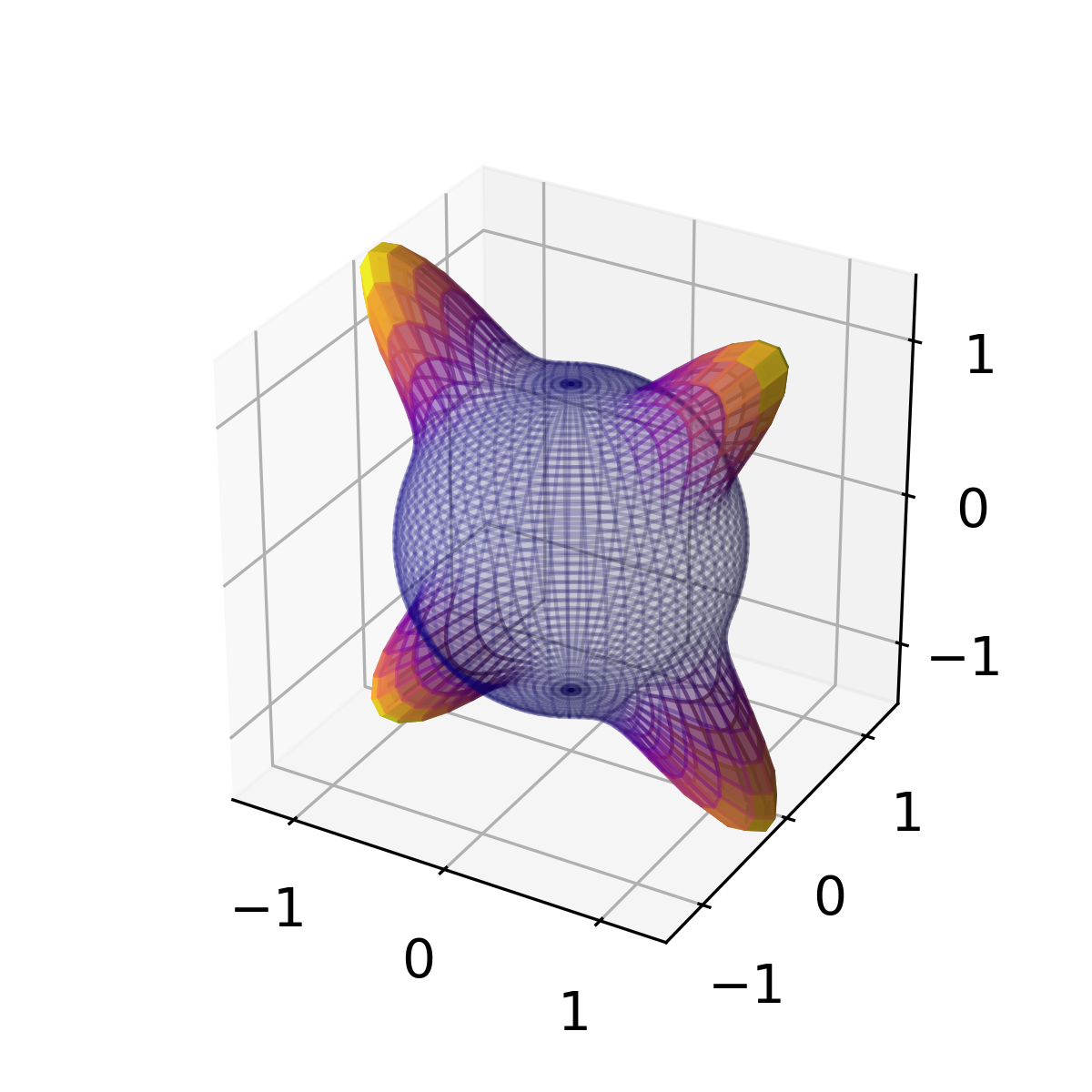}} \\
  40 & A$_1$ & 22521.8451486 & B$_2$ & 22521.8405356&  -2.4 & \adjustbox{valign=m}{\includegraphics[width=2cm]{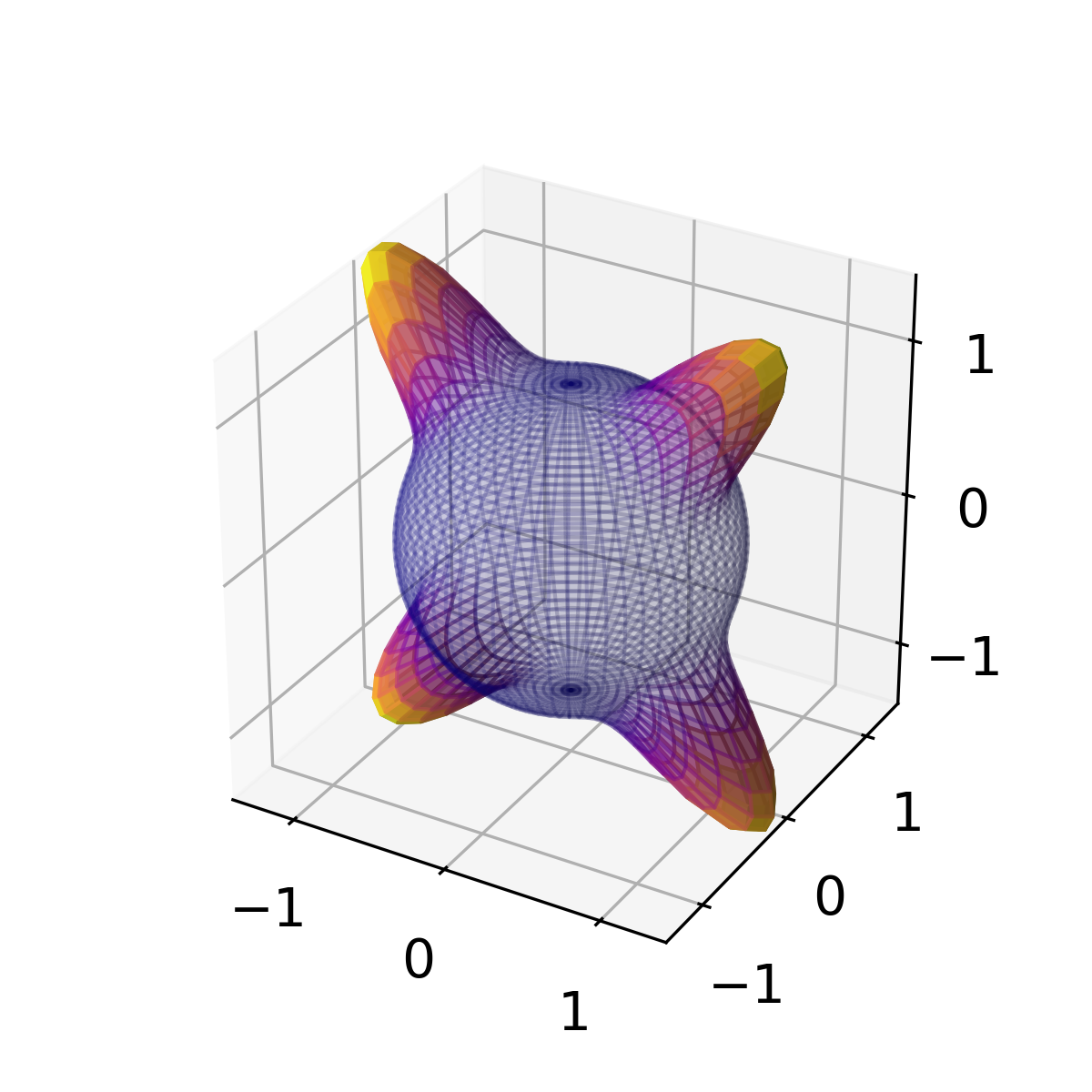}} & \adjustbox{valign=m}{\includegraphics[width=2cm]{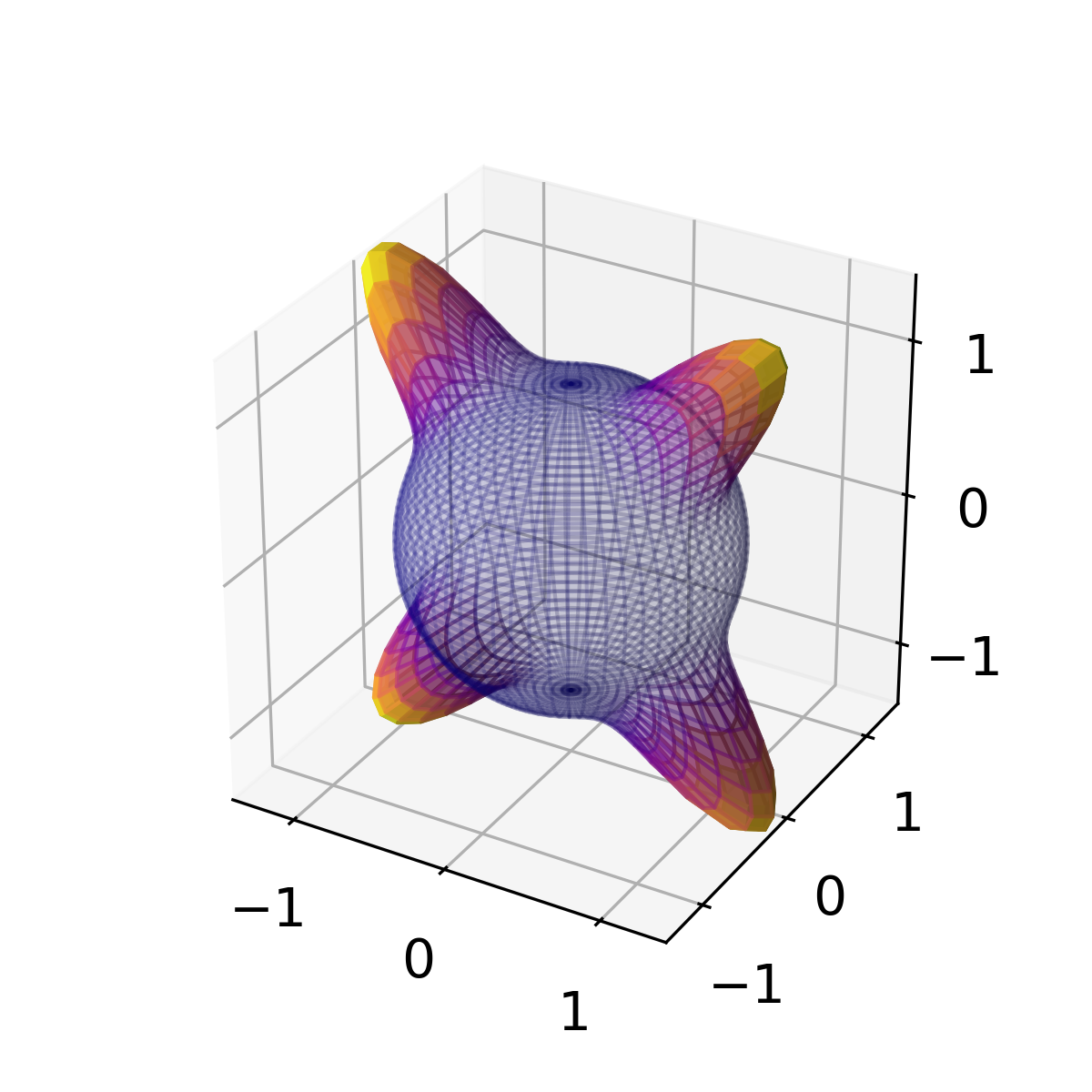}} \\
  40 & A$_1$ & 22847.0931718 & B$_2$ & 22847.0932478&  2.5 &  \adjustbox{valign=m}{\includegraphics[width=2cm]{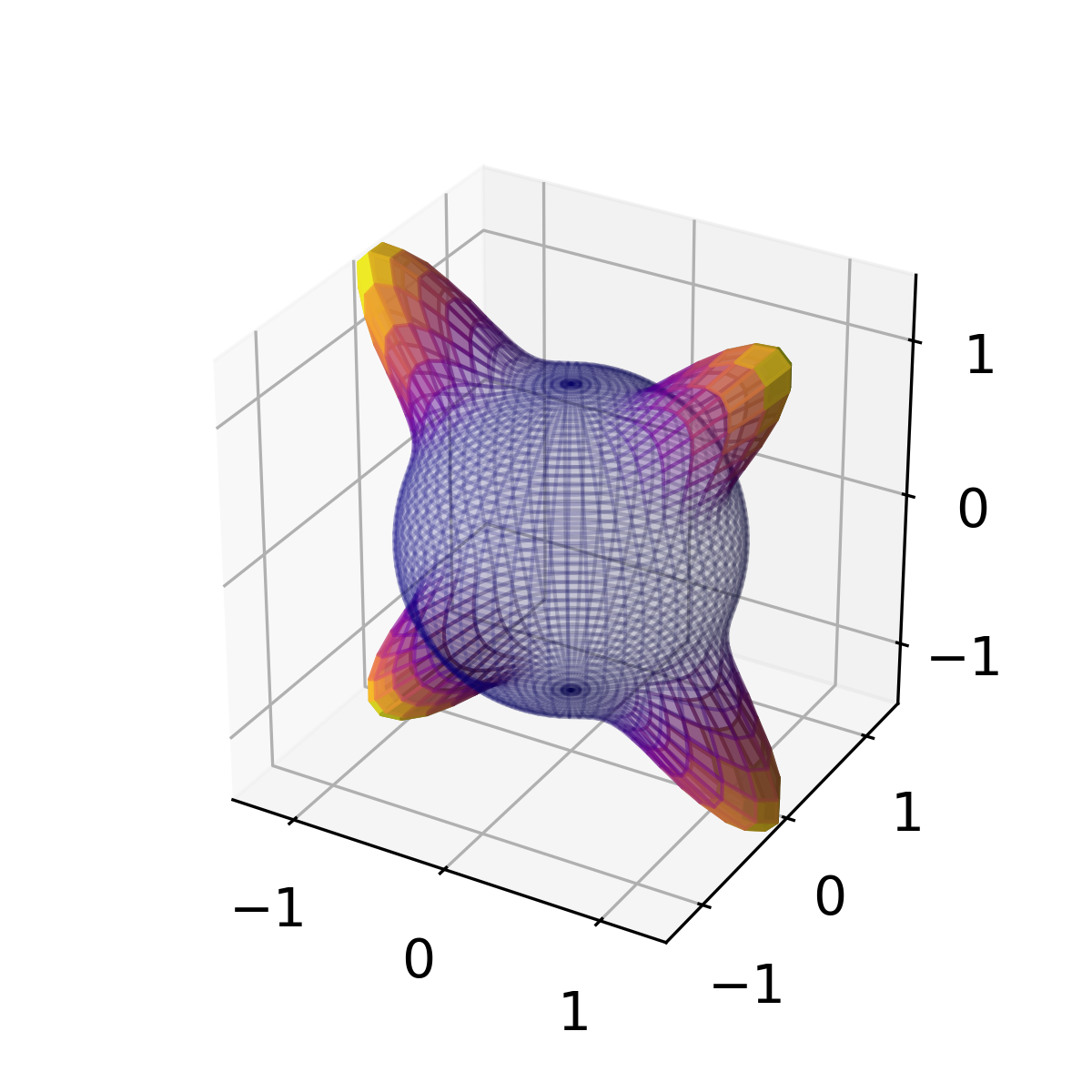}} & \adjustbox{valign=m}{\includegraphics[width=2cm]{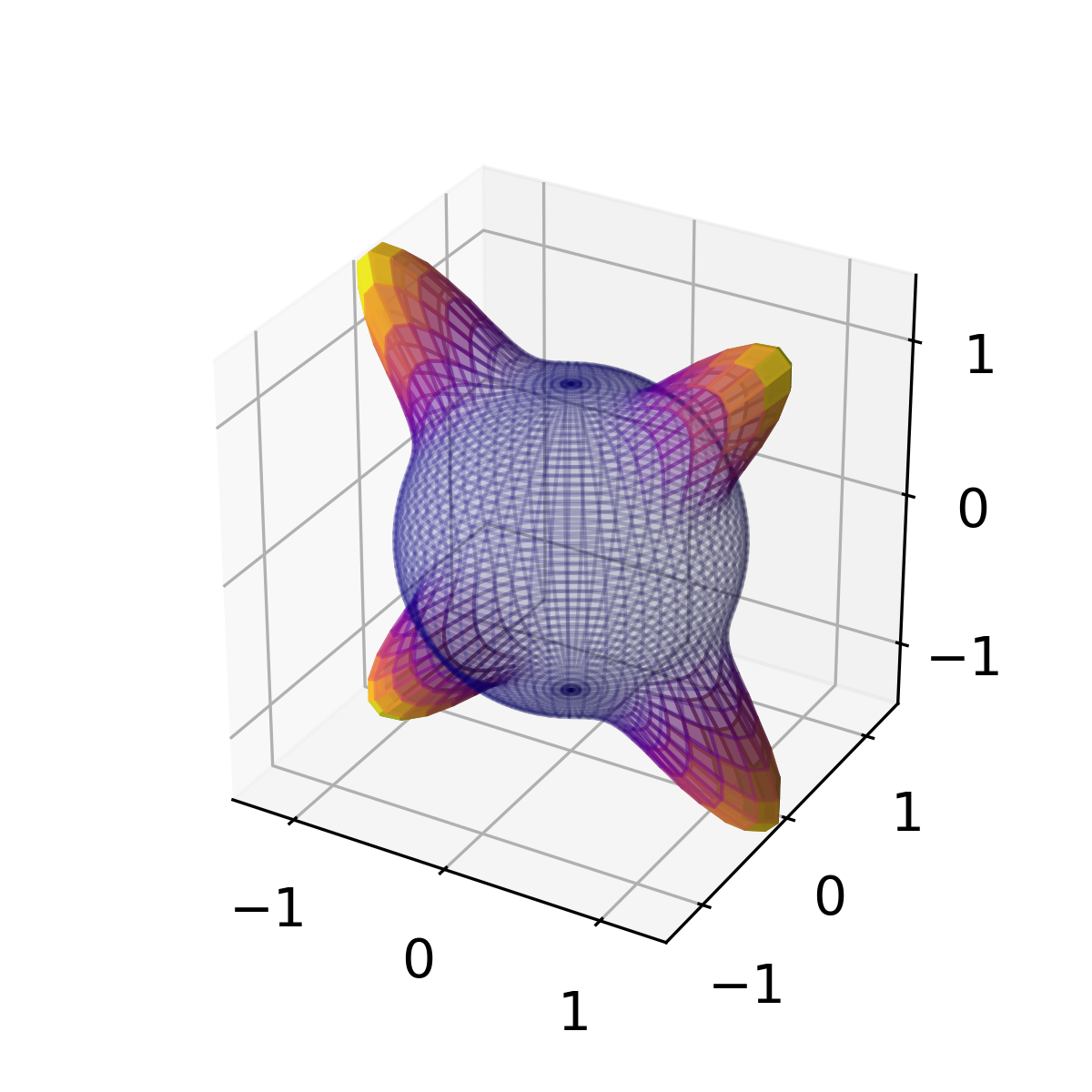}} \\
  45 & A$_1$ & 23717.3037219 & B$_2$ & 23717.3035164&  2.3 &  \adjustbox{valign=m}{\includegraphics[width=2cm]{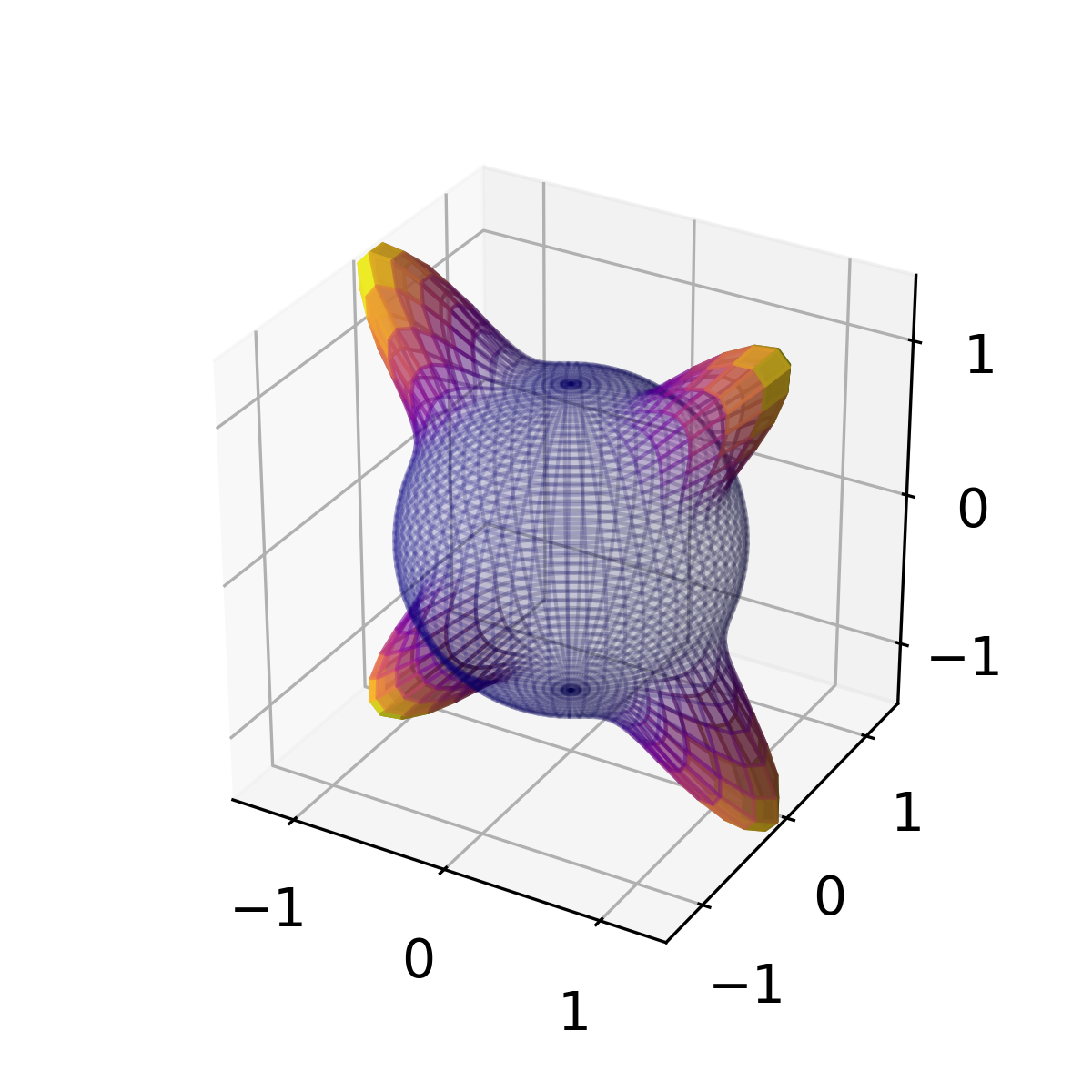}} & \adjustbox{valign=m}{\includegraphics[width=2cm]{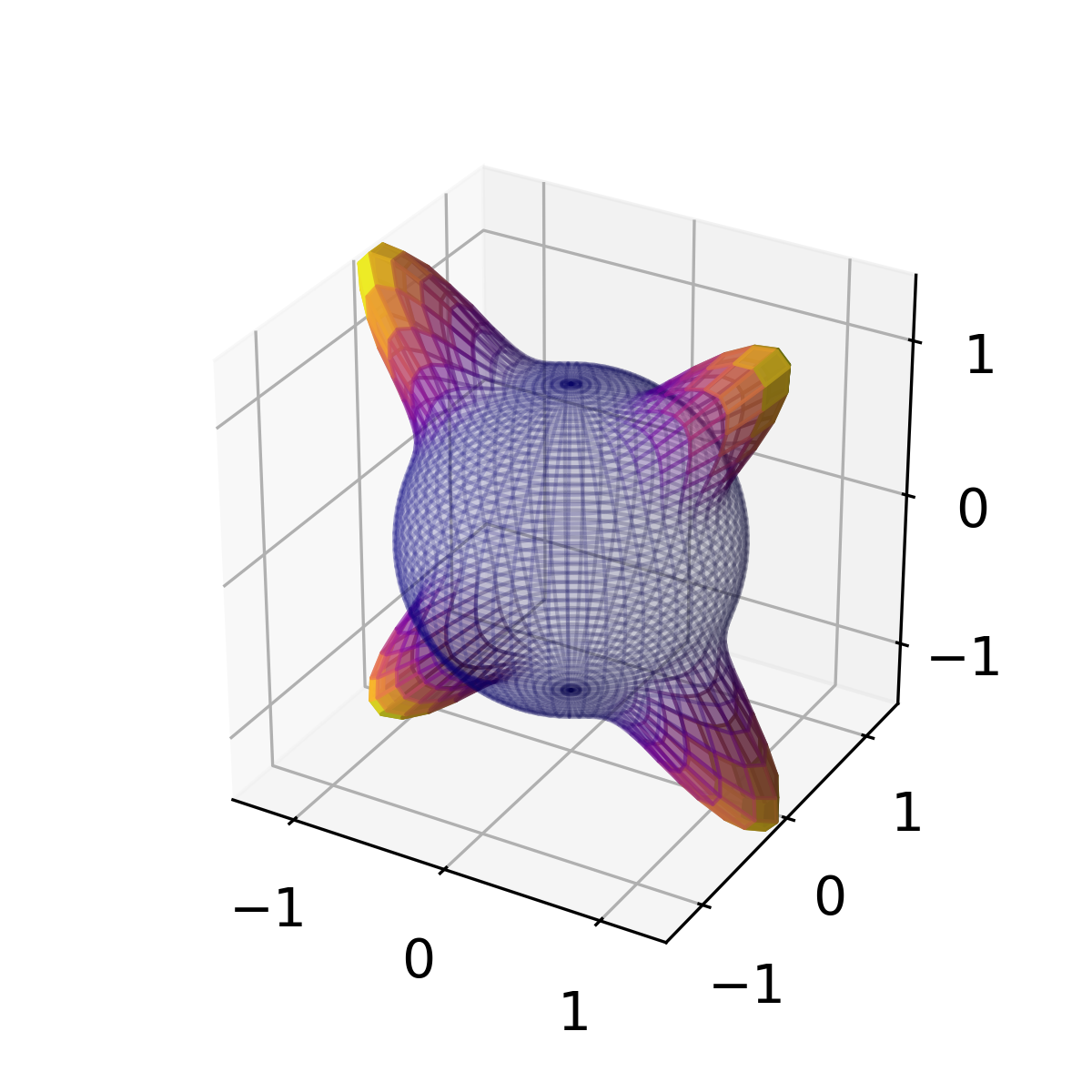}} \\
  45 & A$_1$ & 25779.9032620 & B$_2$ & 25779.9031540&  -2.3&  \adjustbox{valign=m}{\includegraphics[width=2cm]{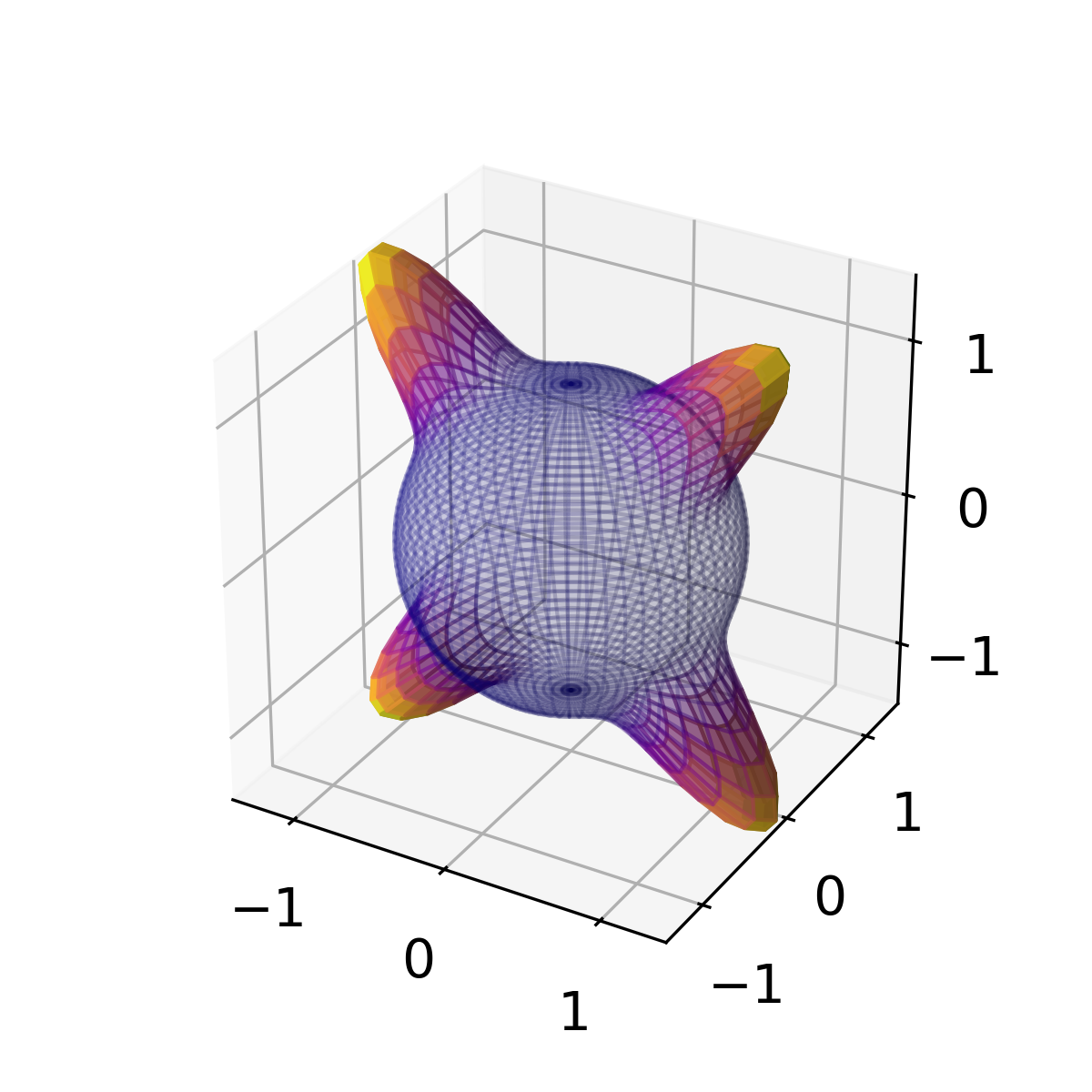}} & \adjustbox{valign=m}{\includegraphics[width=2cm]{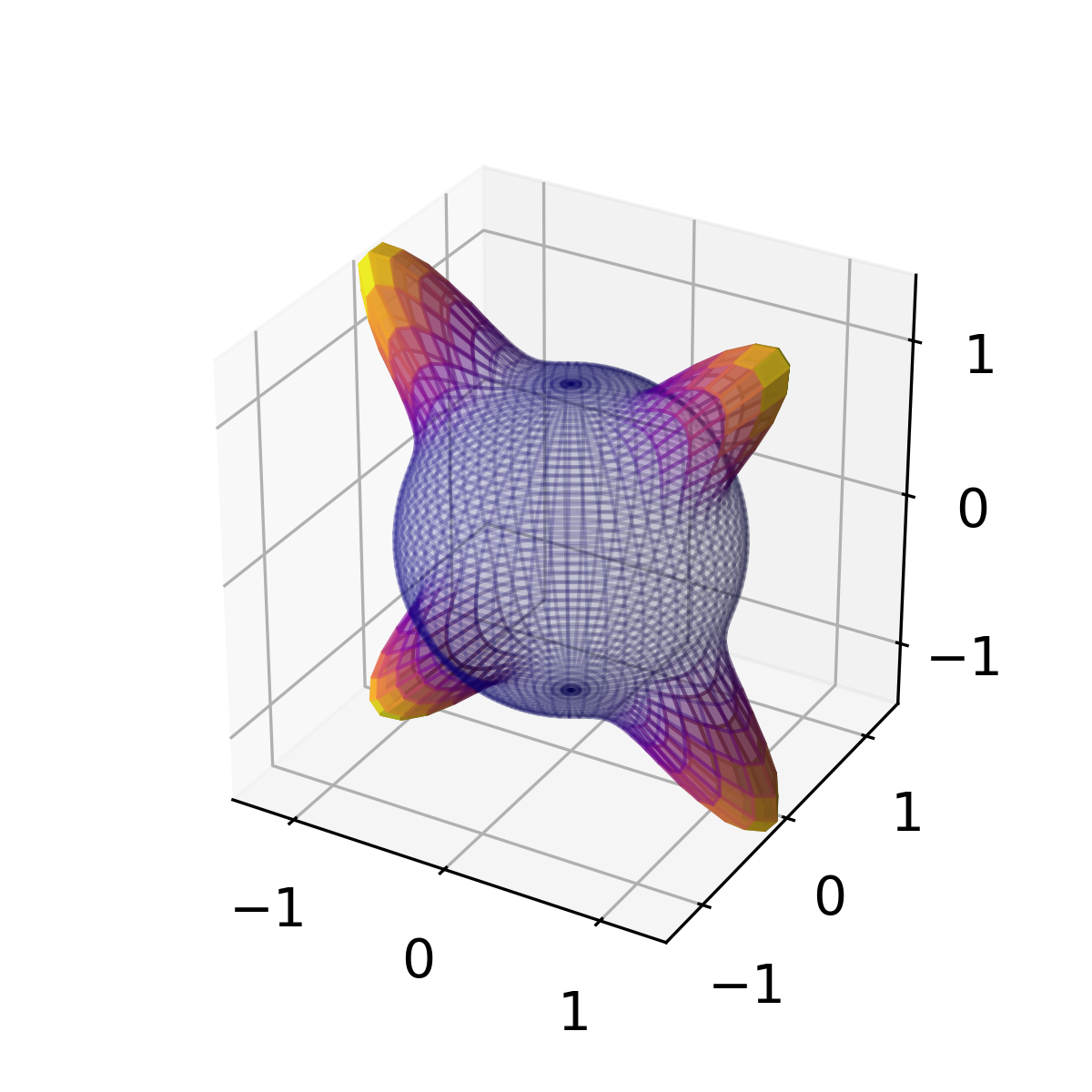}} \\
  45 & A$_1$ & 26220.1563466 & B$_2$ & 26220.1565564&  -2.3&  \adjustbox{valign=m}{\includegraphics[width=2cm]{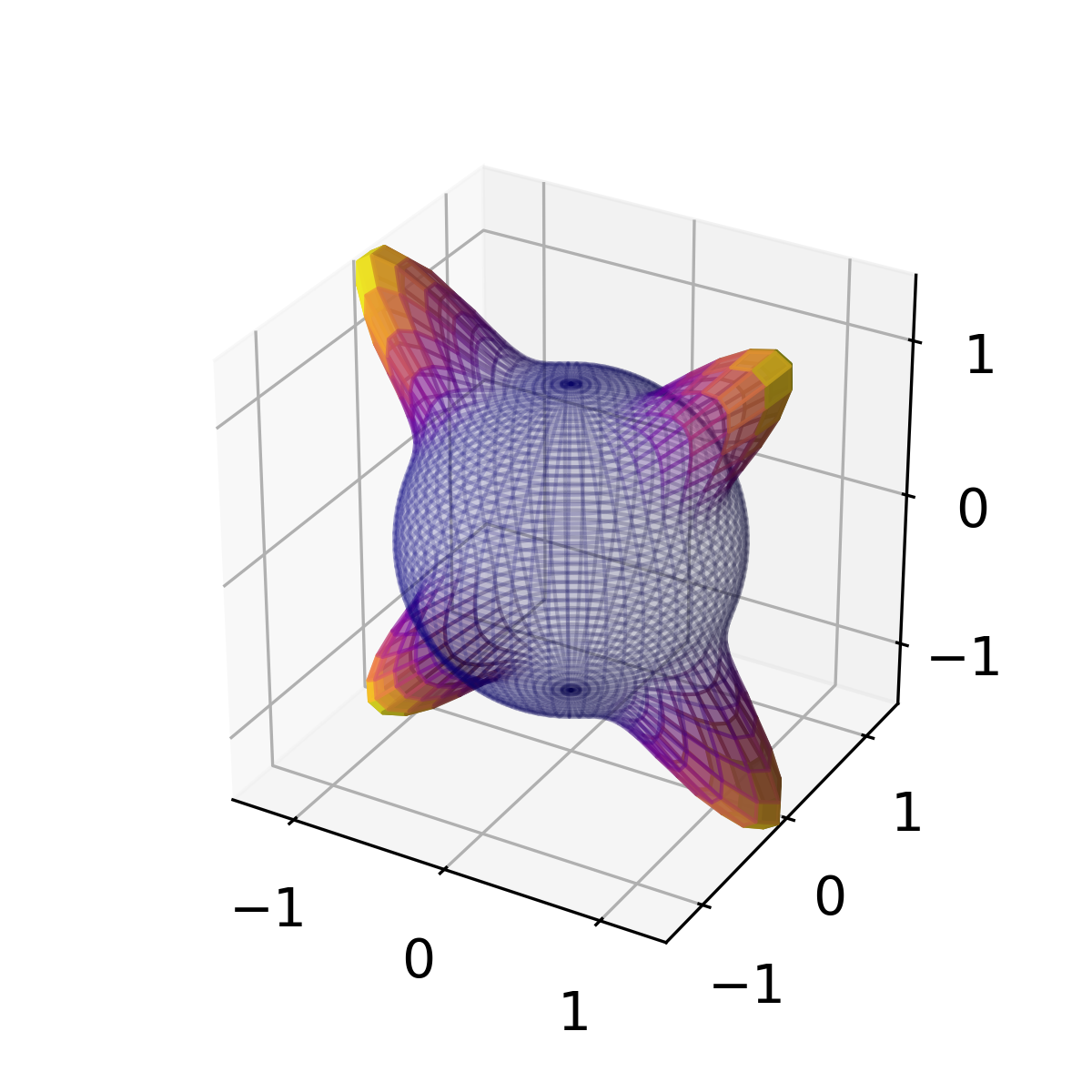}} & \adjustbox{valign=m}{\includegraphics[width=2cm]{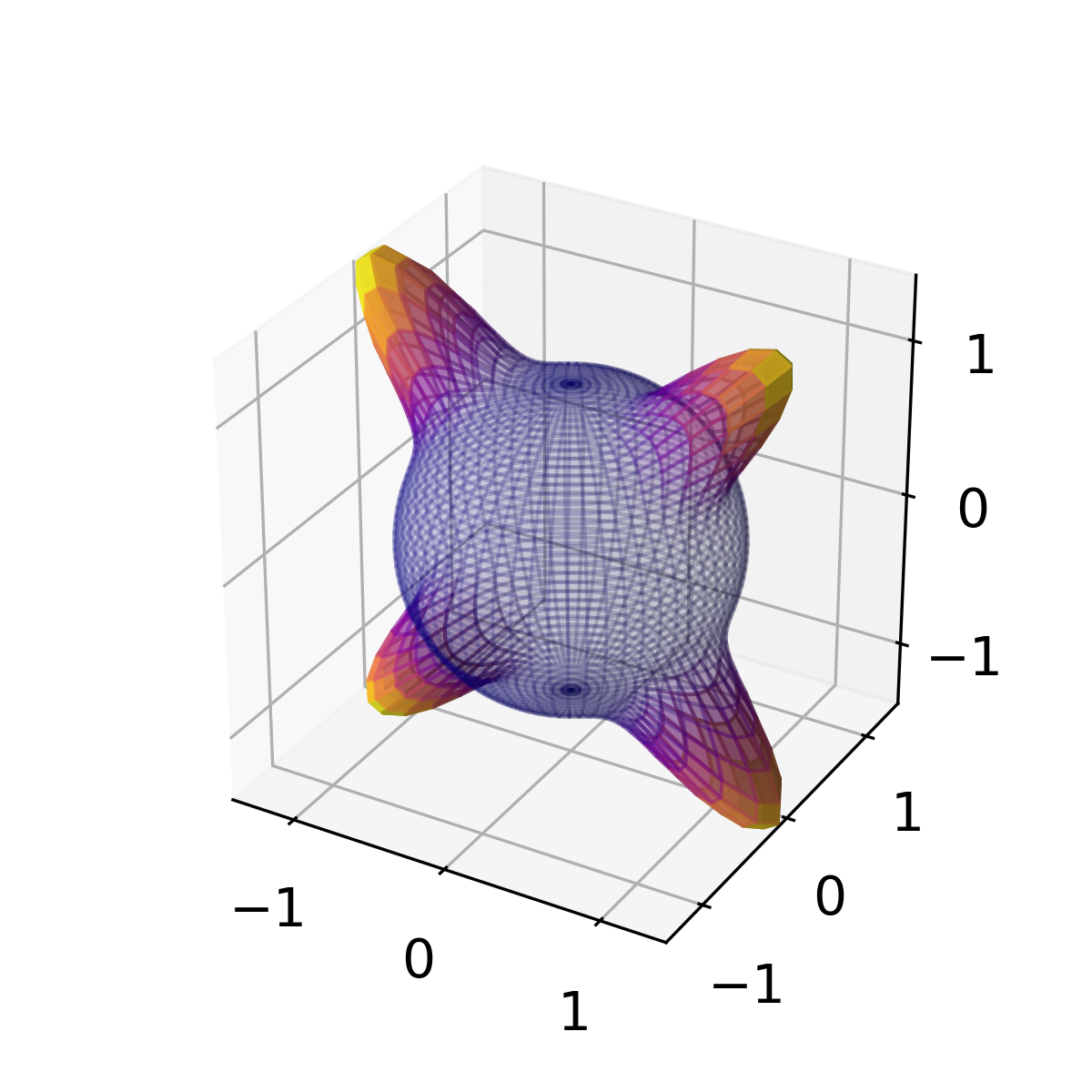}} \\
  50 & A$_1$ & 24708.8188791 & B$_2$ & 24708.8187749&  -2.2&  \adjustbox{valign=m}{\includegraphics[width=2cm]{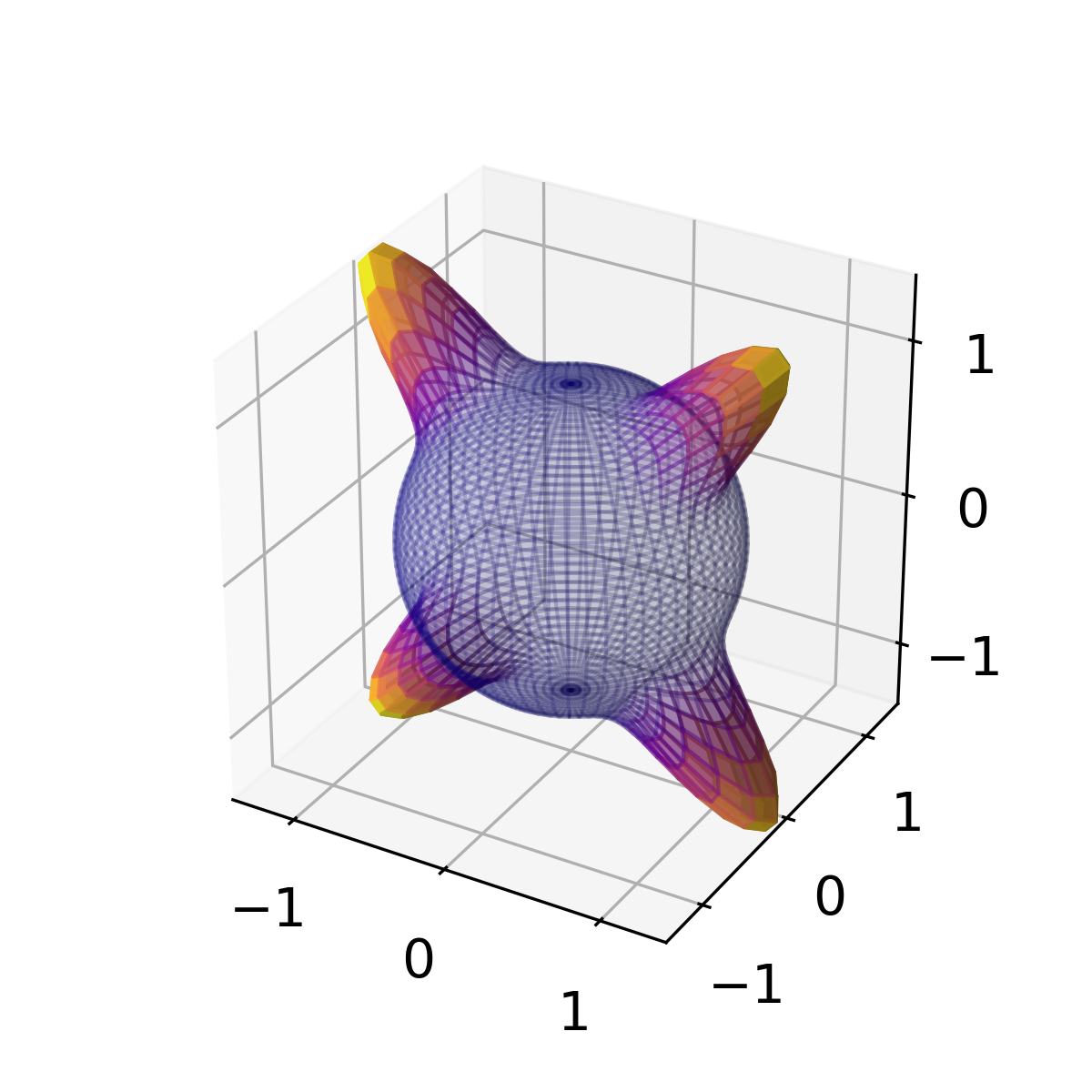}} & \adjustbox{valign=m}{\includegraphics[width=2cm]{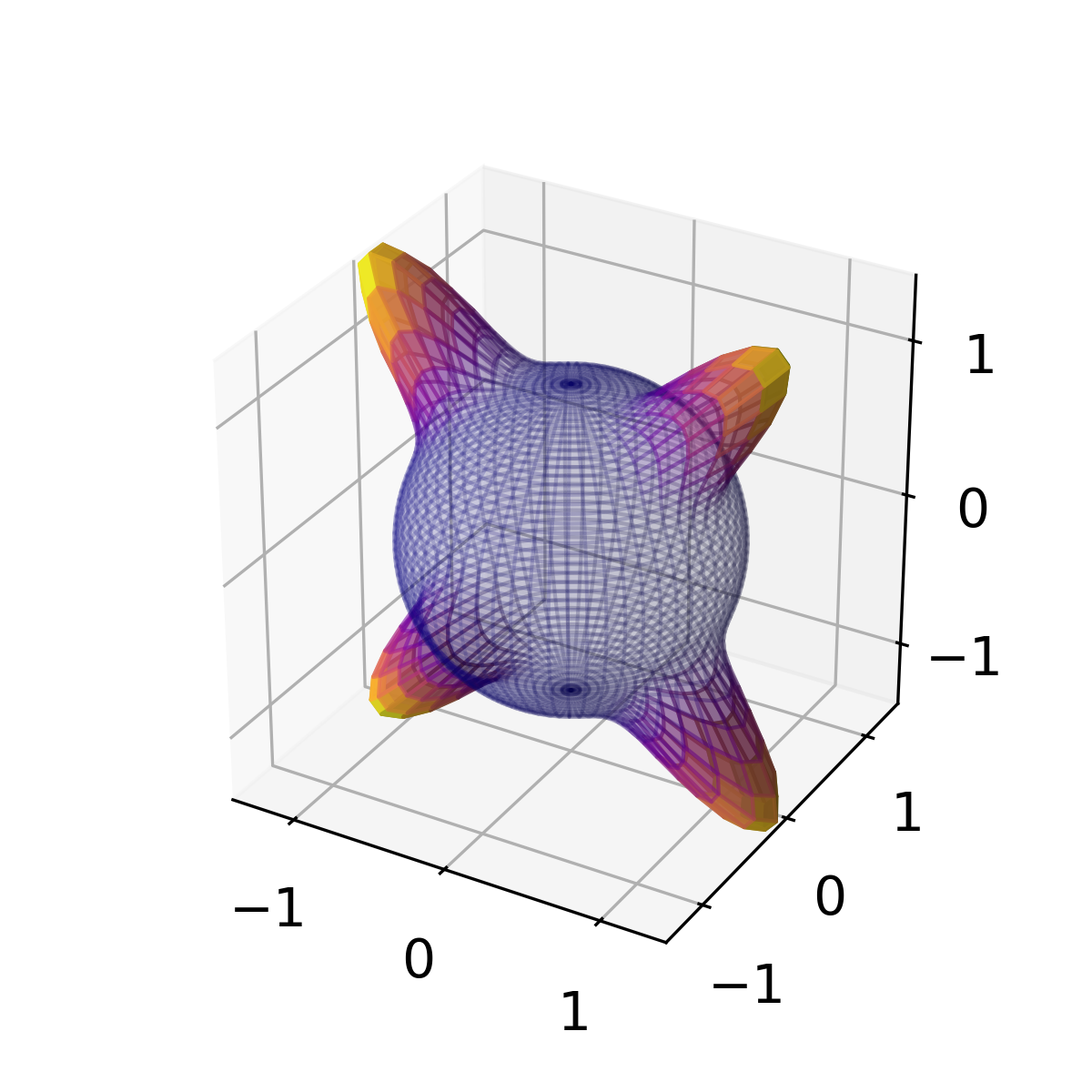}} \\
  50 & A$_1$ & 26642.4819200 & B$_2$ & 26642.4816321&  2.2 &  \adjustbox{valign=m}{\includegraphics[width=2cm]{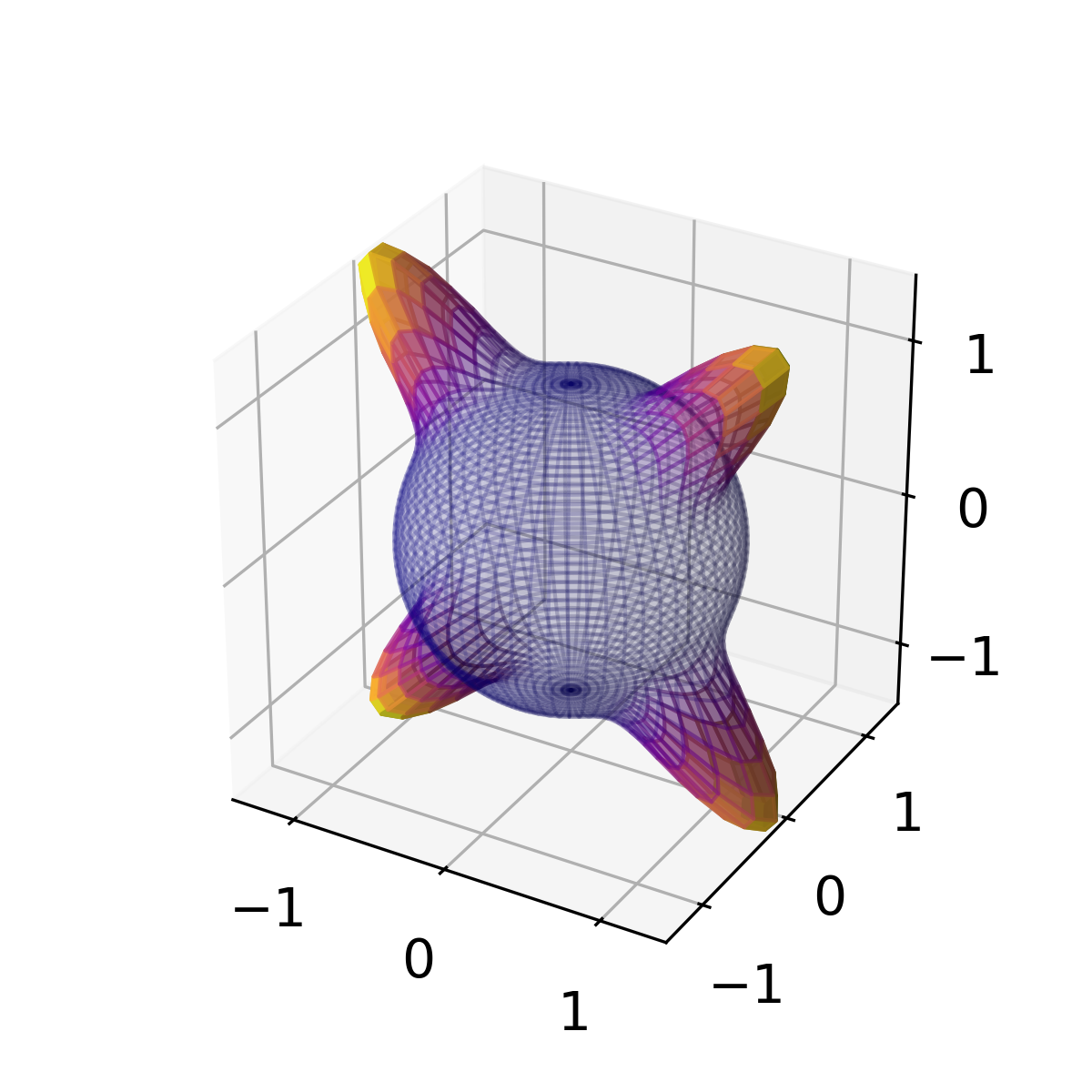}} & \adjustbox{valign=m}{\includegraphics[width=2cm]{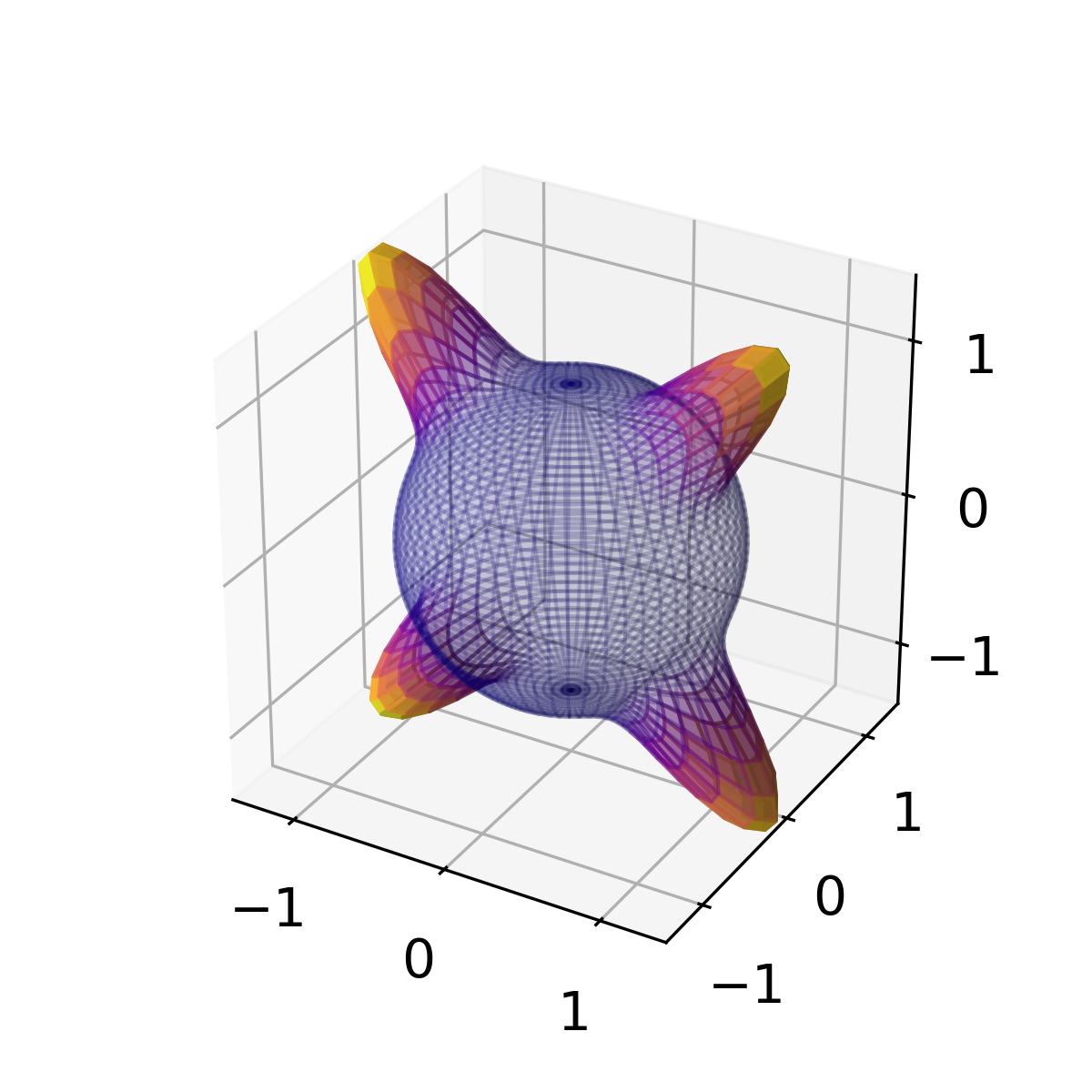}} \\
  50 & A$_1$ & 27304.5875573 & B$_2$ & 27304.5875656&  -2.2&  \adjustbox{valign=m}{\includegraphics[width=2cm]{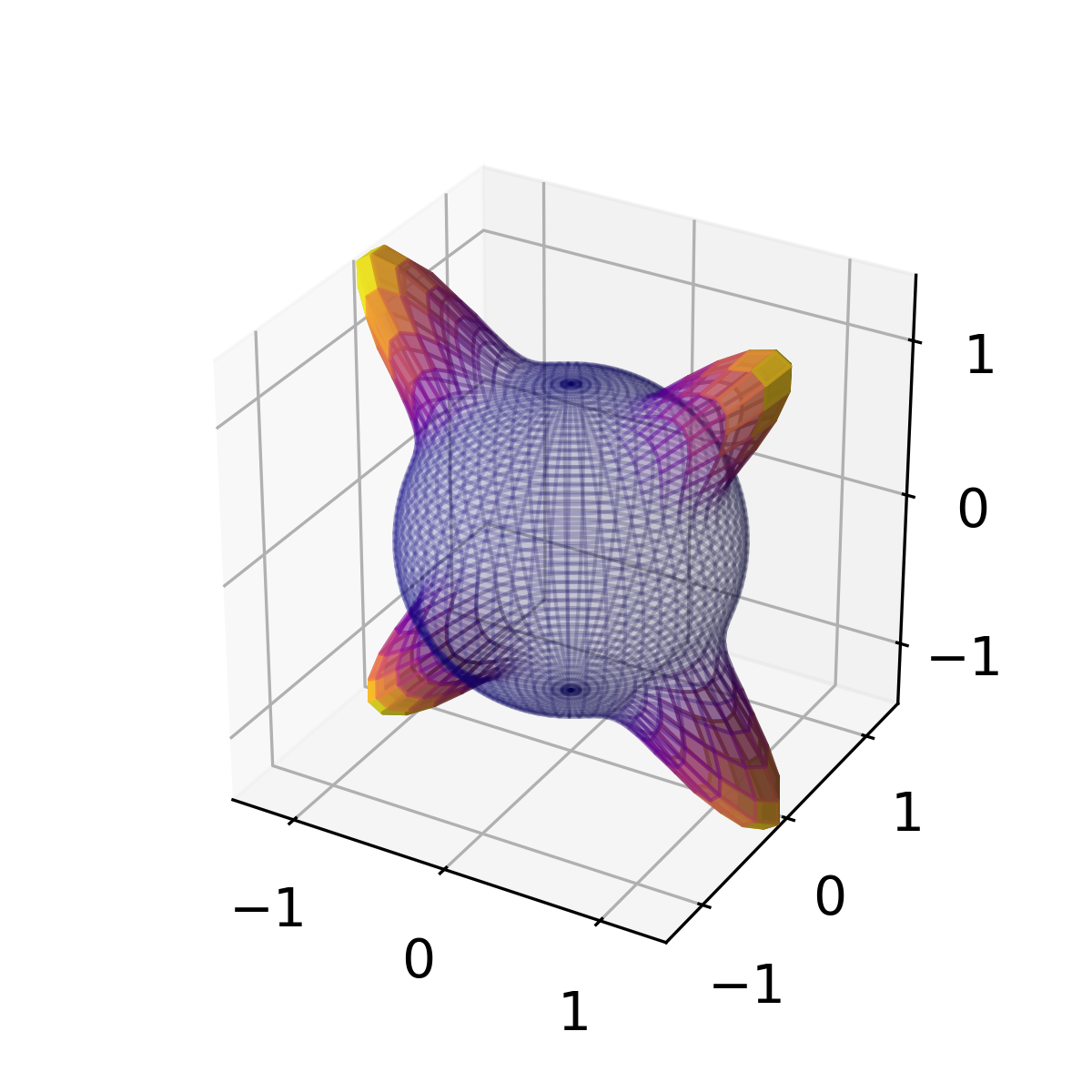}} & \adjustbox{valign=m}{\includegraphics[width=2cm]{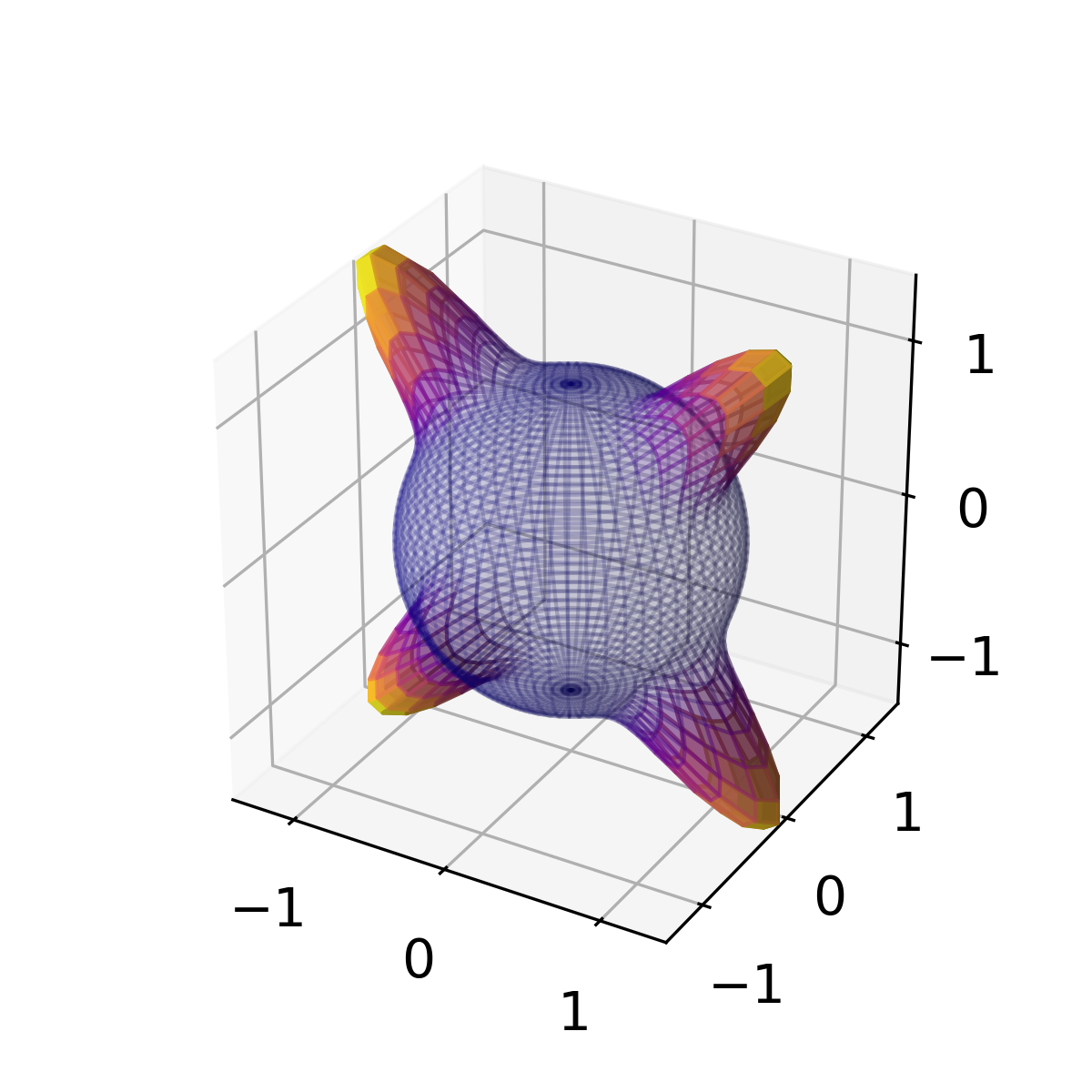}} \\
  55 & A$_1$ & 28457.3454107 & B$_2$ & 28457.3454199&  -2.1&  \adjustbox{valign=m}{\includegraphics[width=2cm]{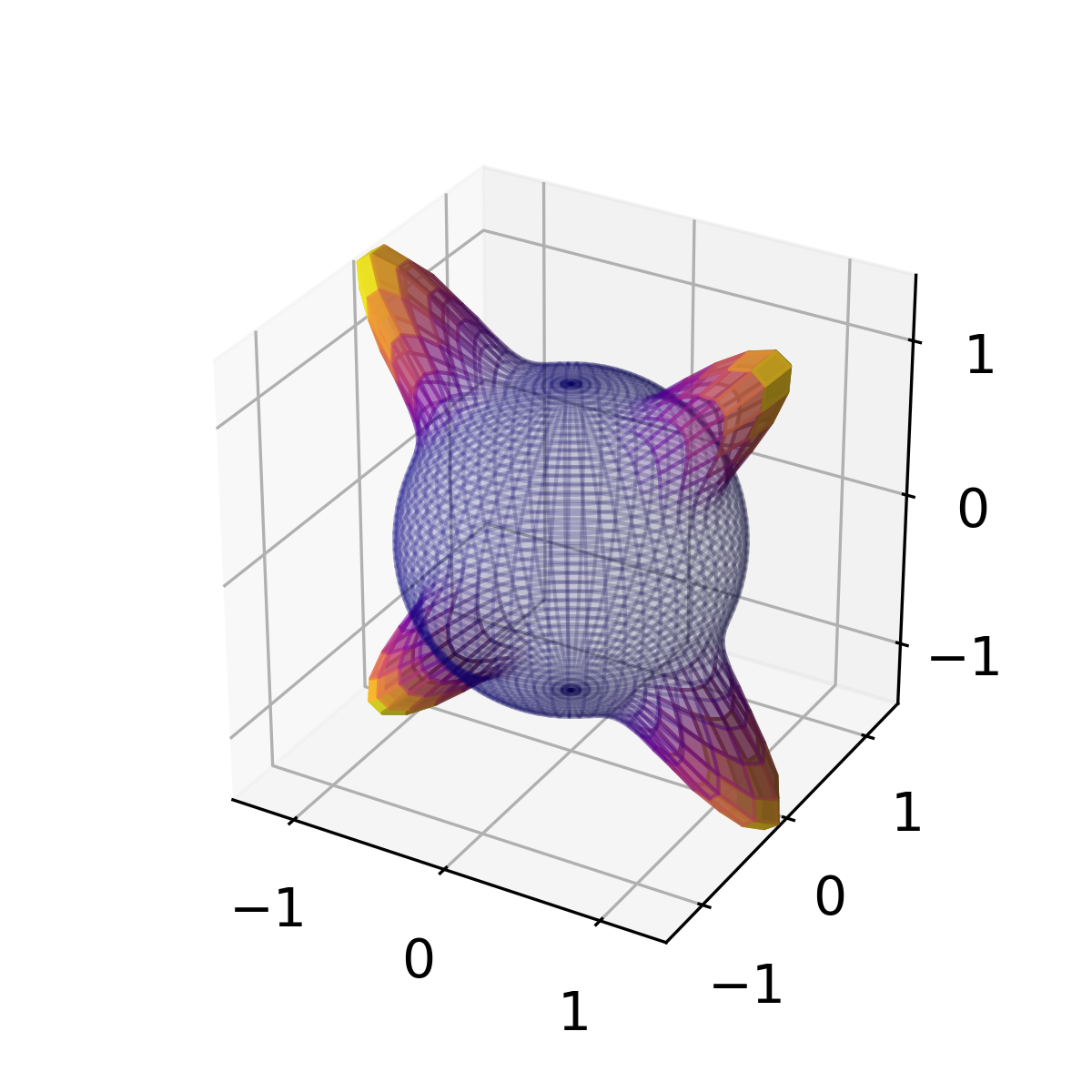}} & \adjustbox{valign=m}{\includegraphics[width=2cm]{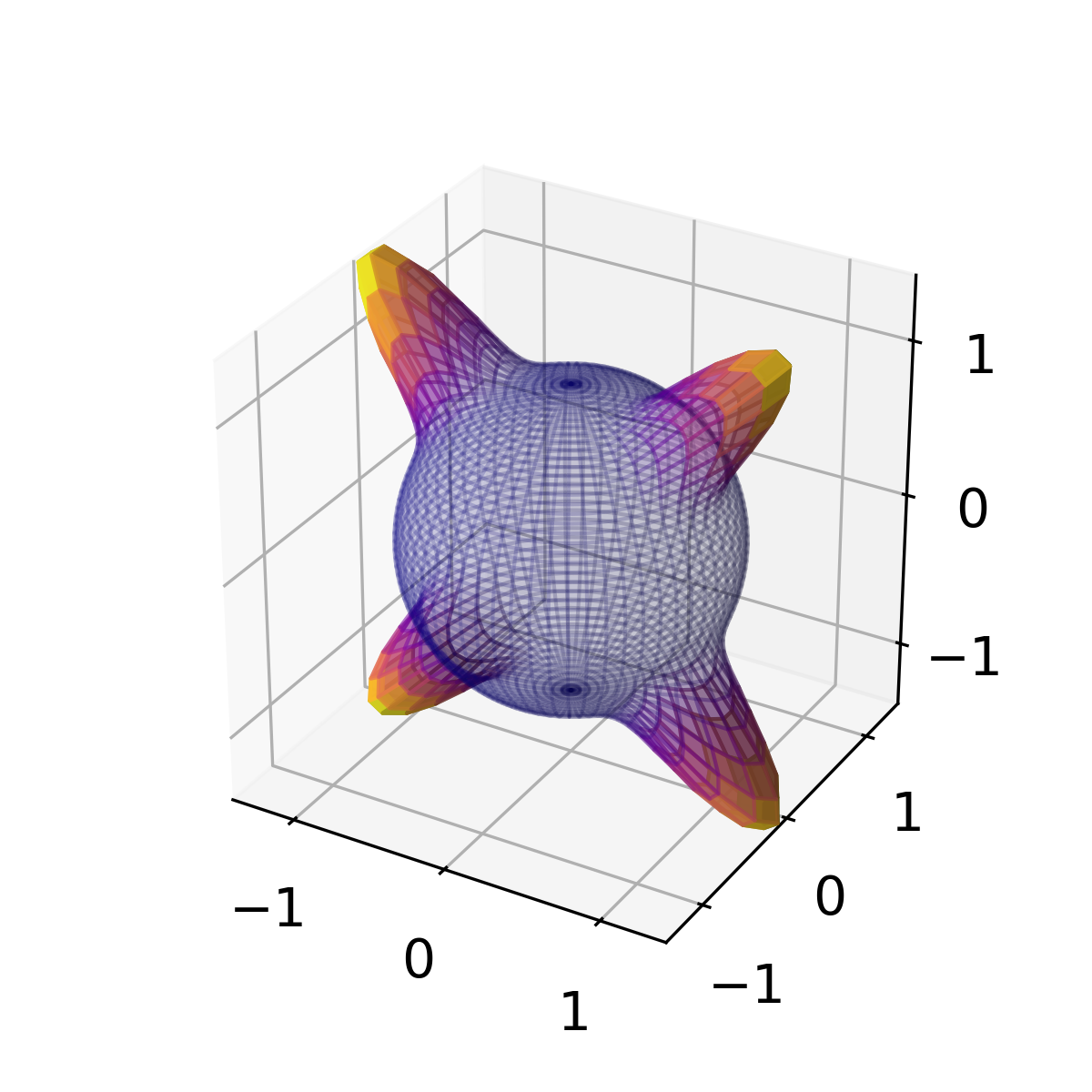}} \\
  55 & A$_1$ & 30244.5163667 & B$_2$ & 30244.5165271&  -2.1&  \adjustbox{valign=m}{\includegraphics[width=2cm]{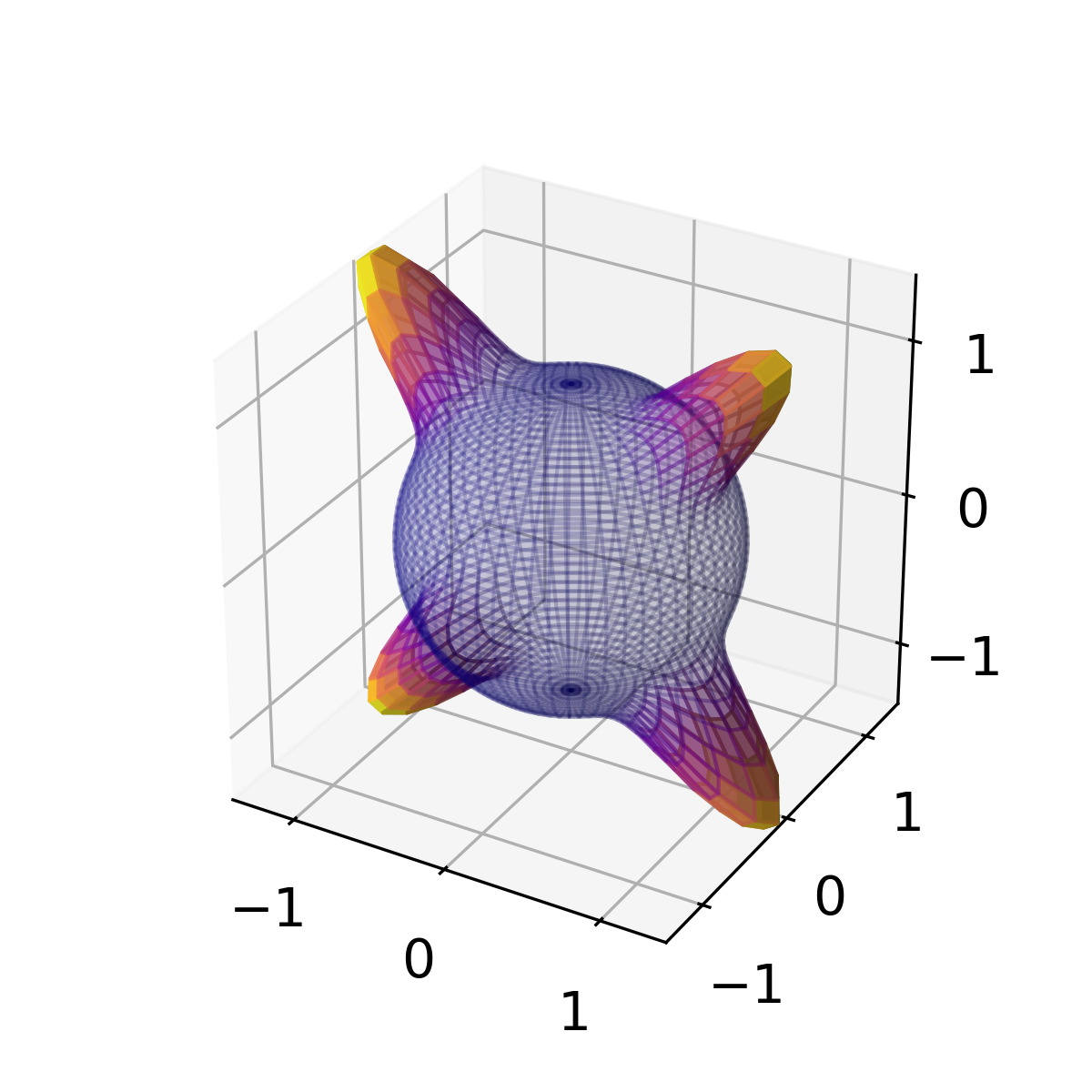}} & \adjustbox{valign=m}{\includegraphics[width=2cm]{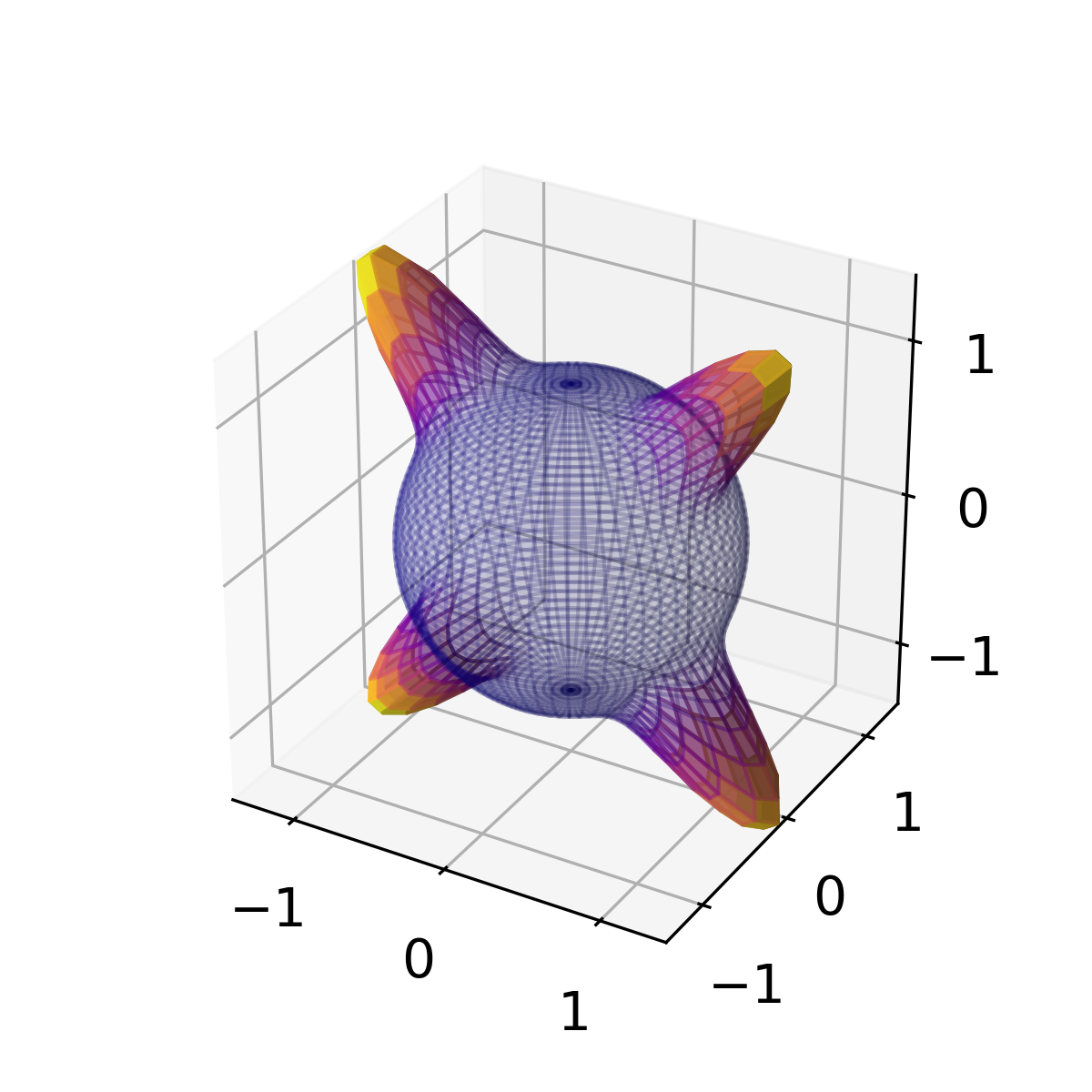}} \\
  55 & A$_1$ & 31053.3626499 & B$_2$ & 31053.3625292&  -2.1&  \adjustbox{valign=m}{\includegraphics[width=2cm]{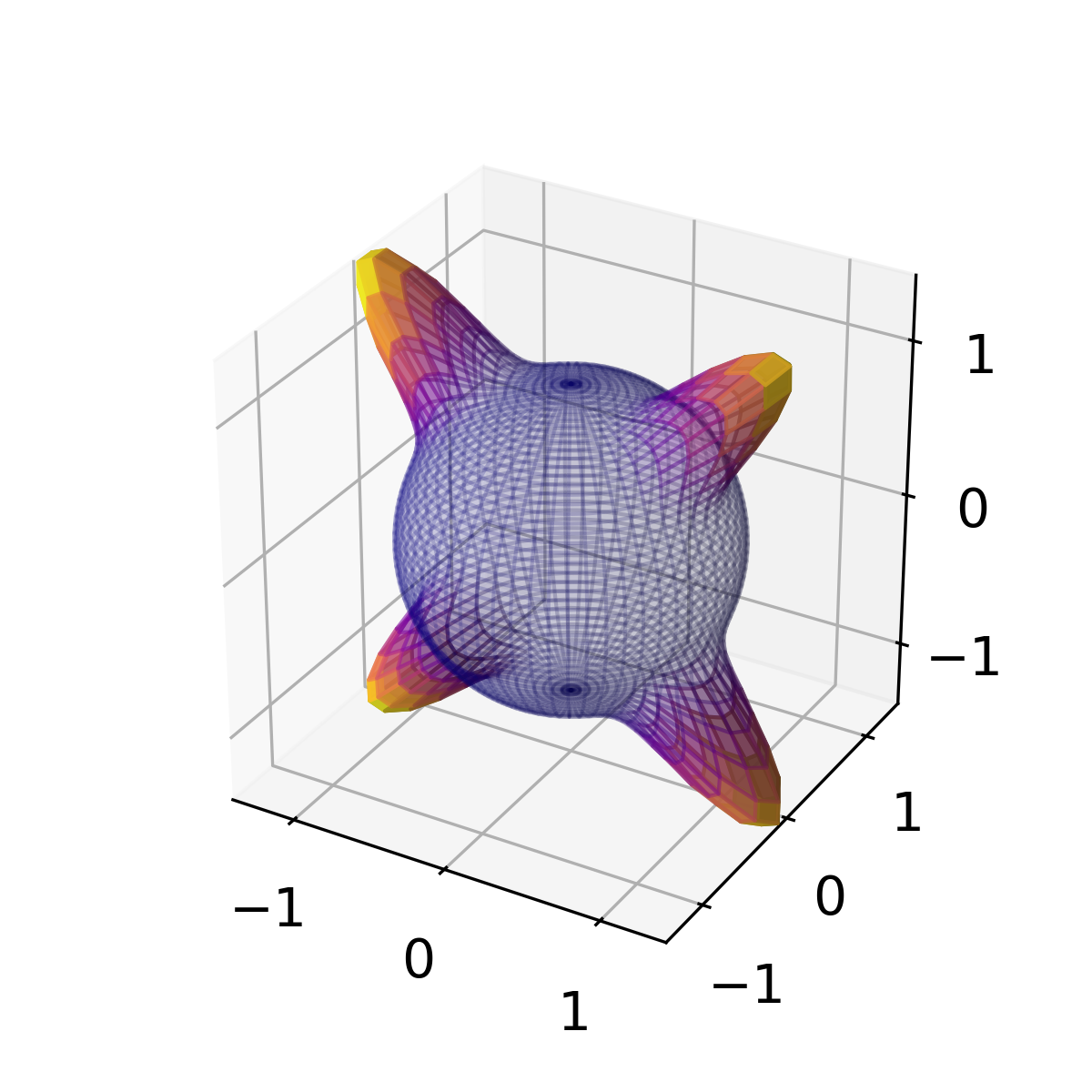}} & \adjustbox{valign=m}{\includegraphics[width=2cm]{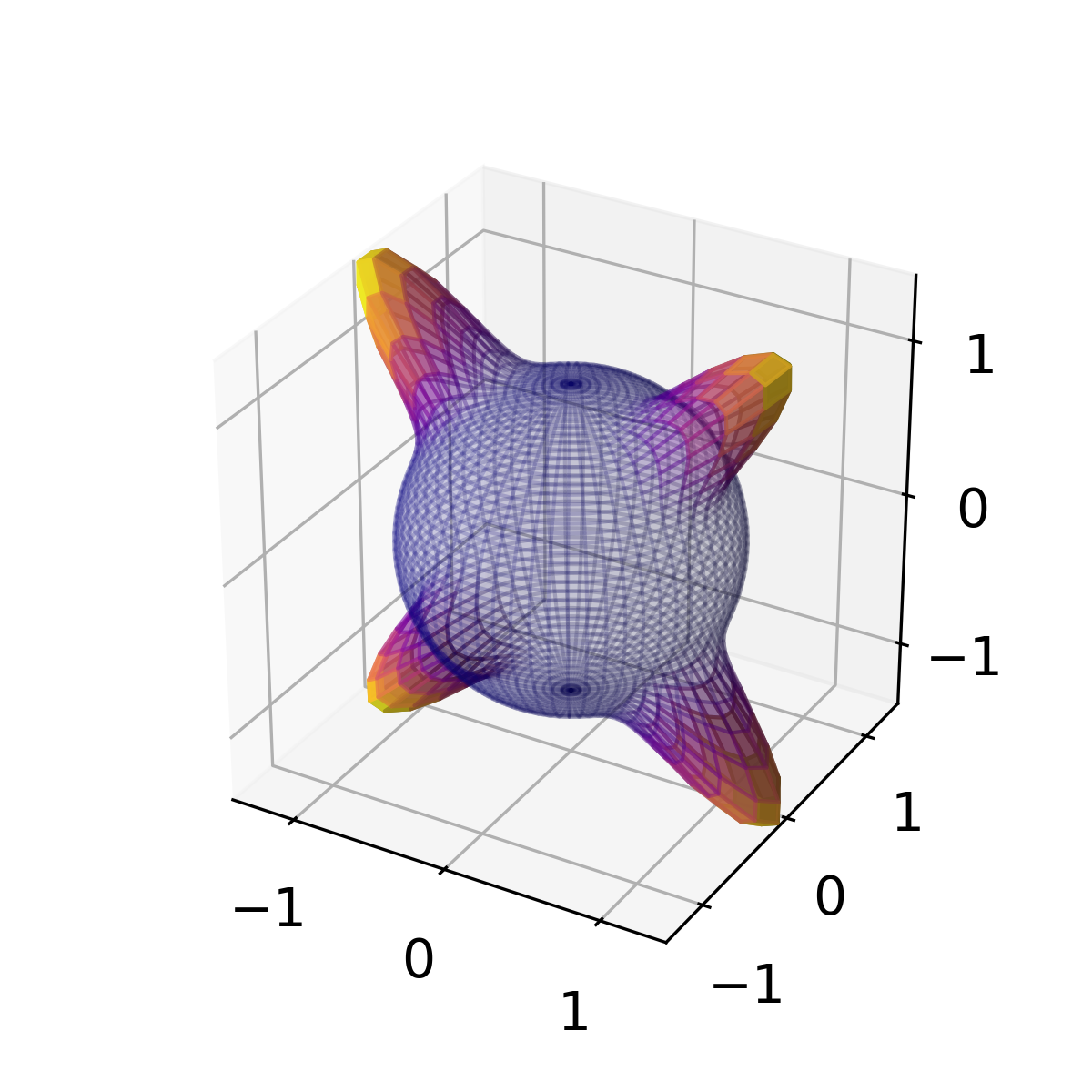}} \\
  60 & A$_1$ & 32309.7953093 & B$_2$ & 32309.7953093&  2.1 &  \adjustbox{valign=m}{\includegraphics[width=2cm]{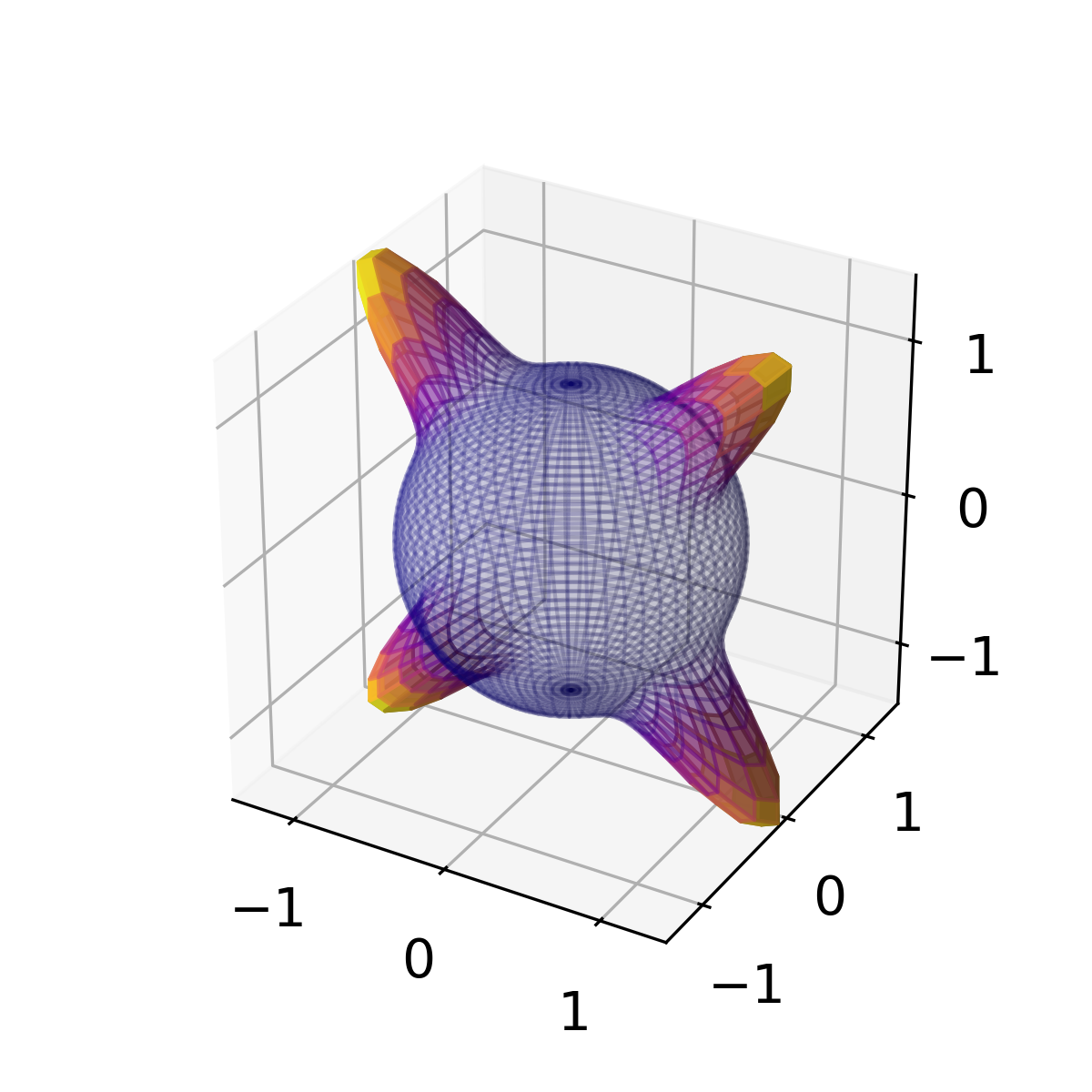}} & \adjustbox{valign=m}{\includegraphics[width=2cm]{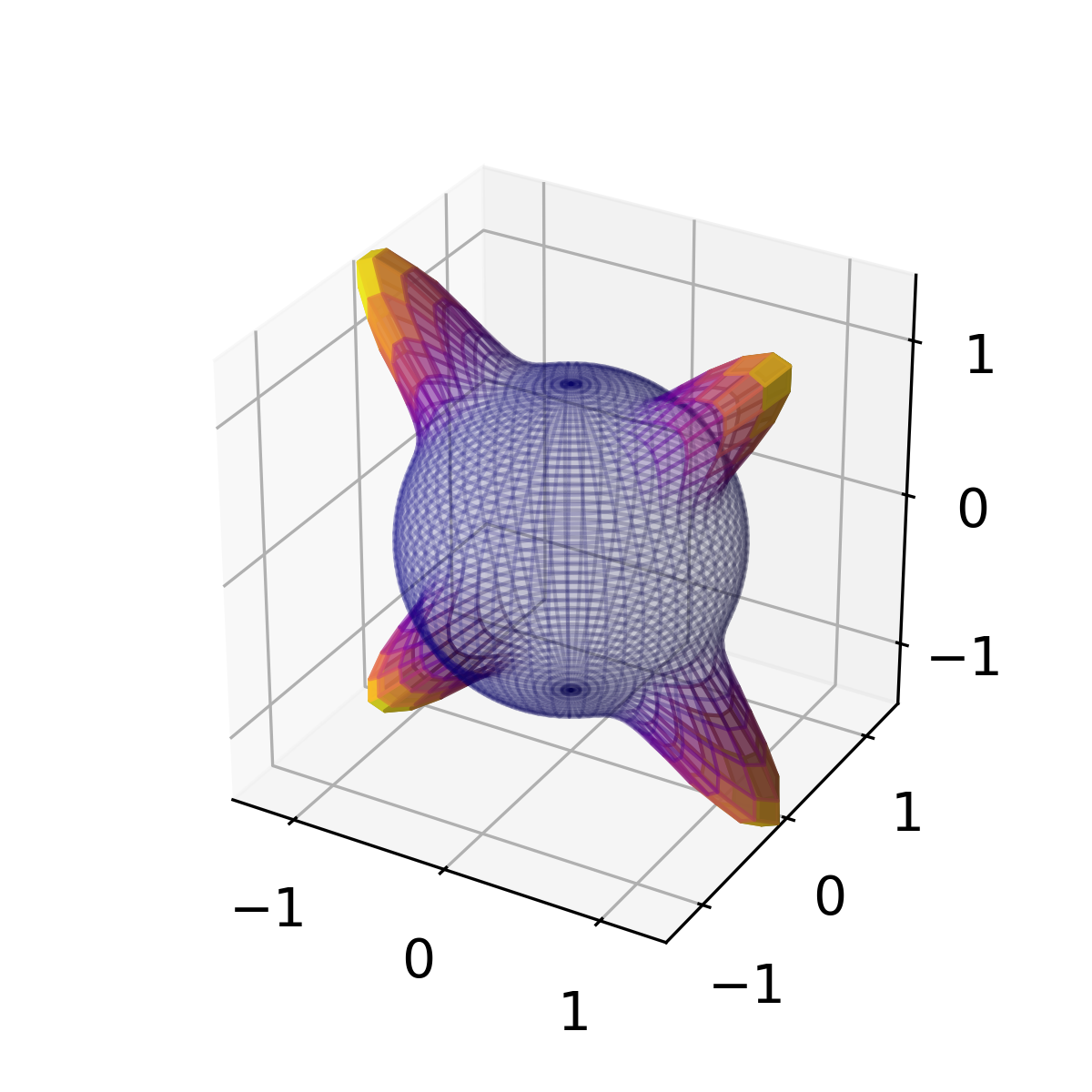}} \\
  60 & A$_1$ & 33735.3399384 & B$_2$ & 33735.3399365&  2.0 &  \adjustbox{valign=m}{\includegraphics[width=2cm]{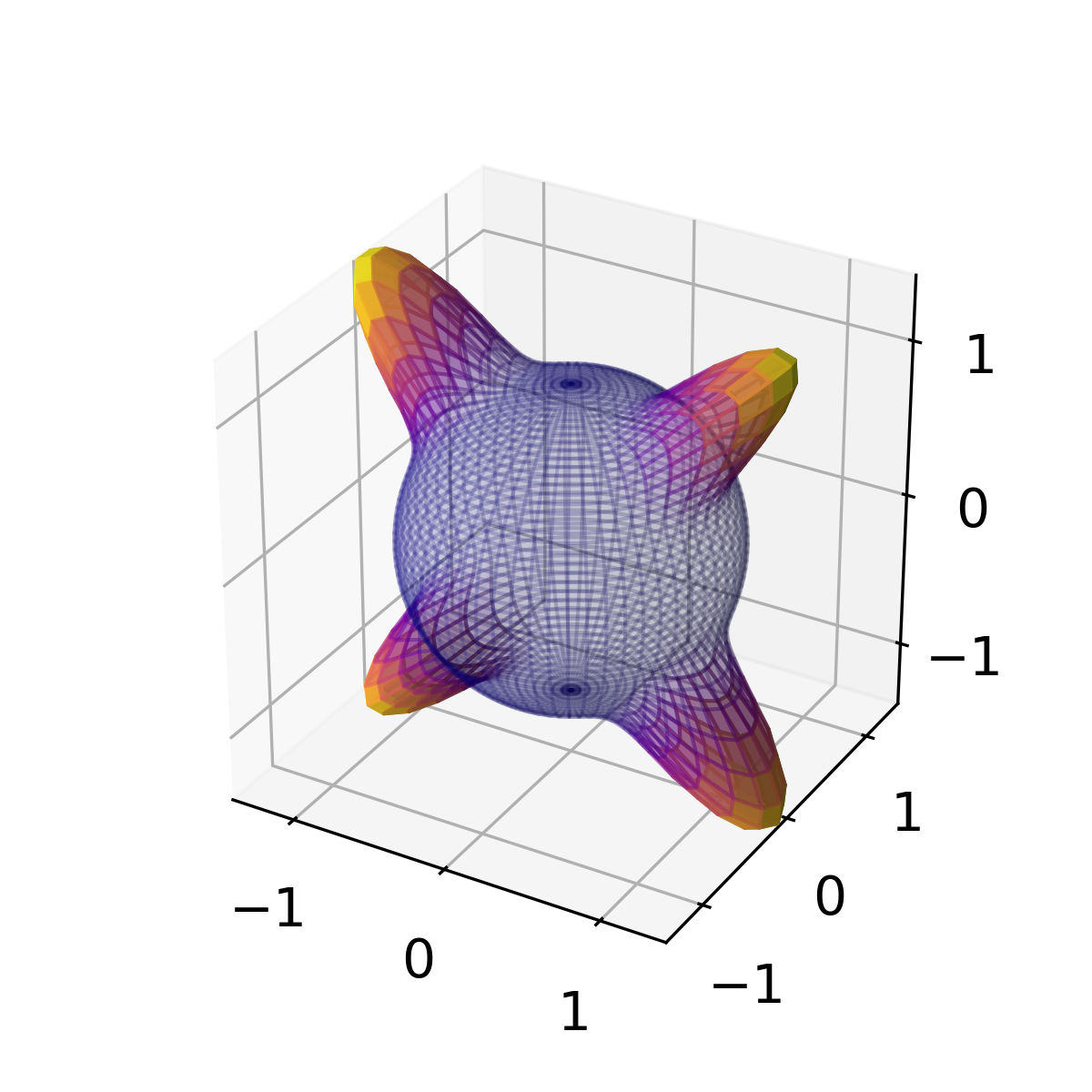}} & \adjustbox{valign=m}{\includegraphics[width=2cm]{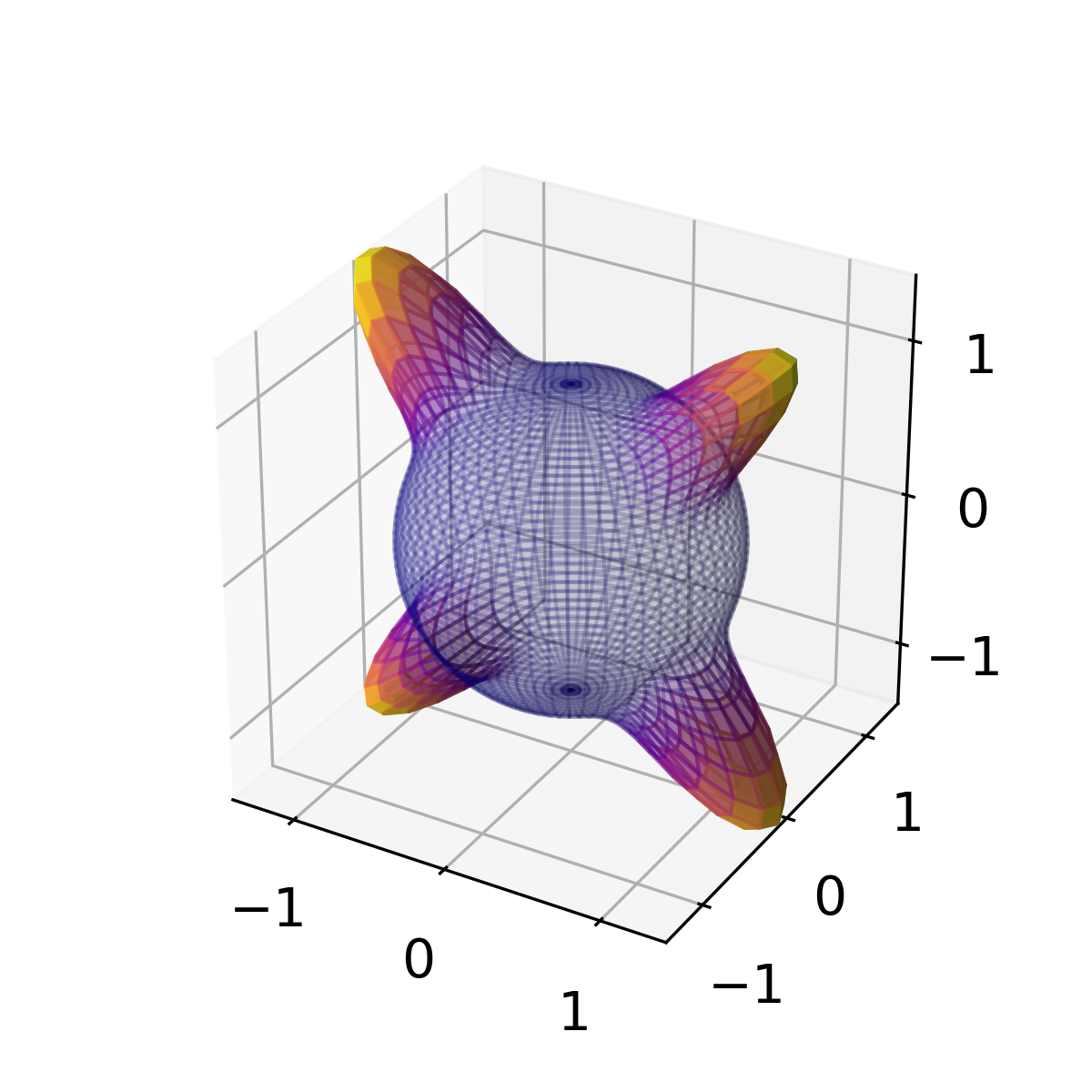}} \\
  60 & A$_1$ & 33933.6589073 & B$_2$ & 33933.6588567&  -2.1&  \adjustbox{valign=m}{\includegraphics[width=2cm]{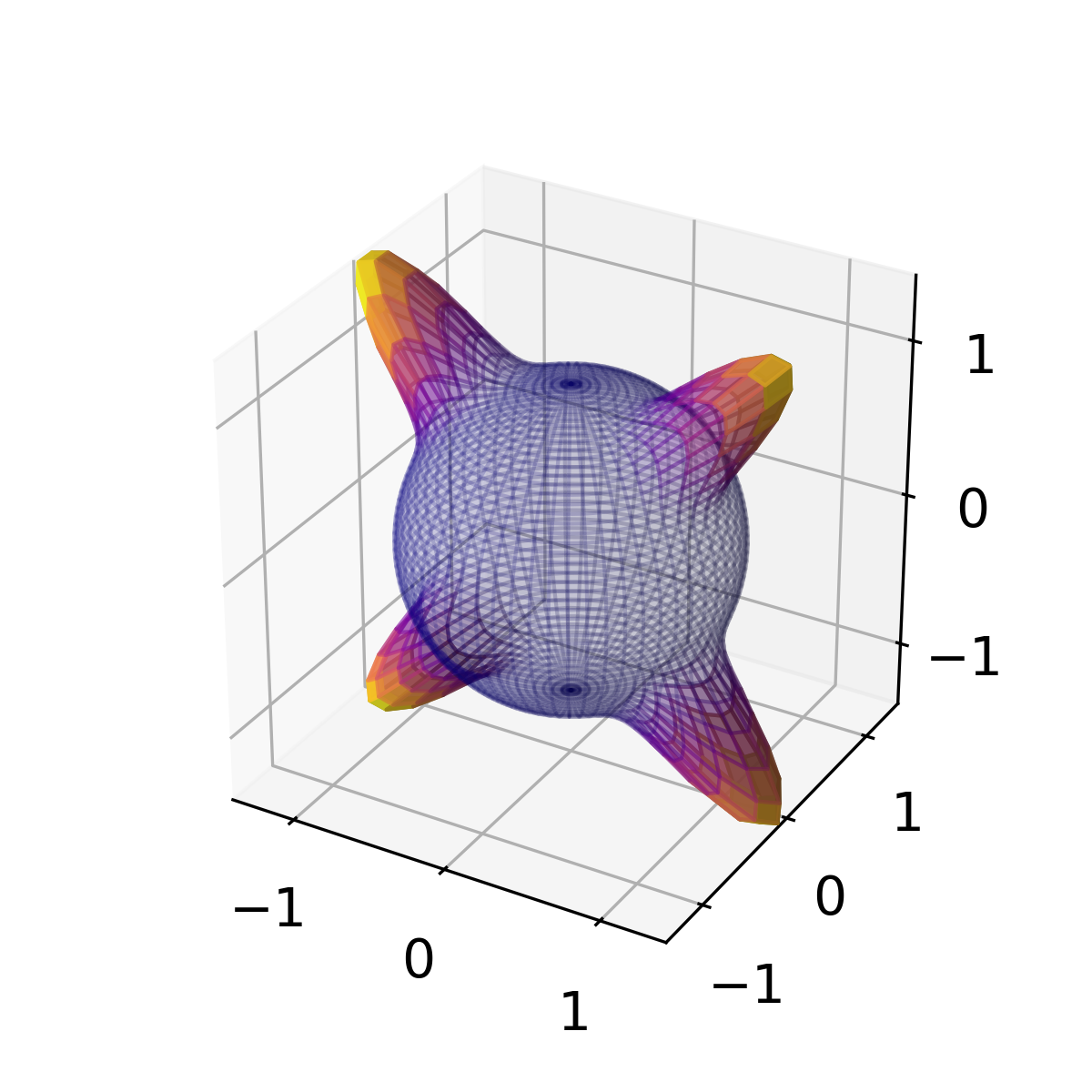}} & \adjustbox{valign=m}{\includegraphics[width=2cm]{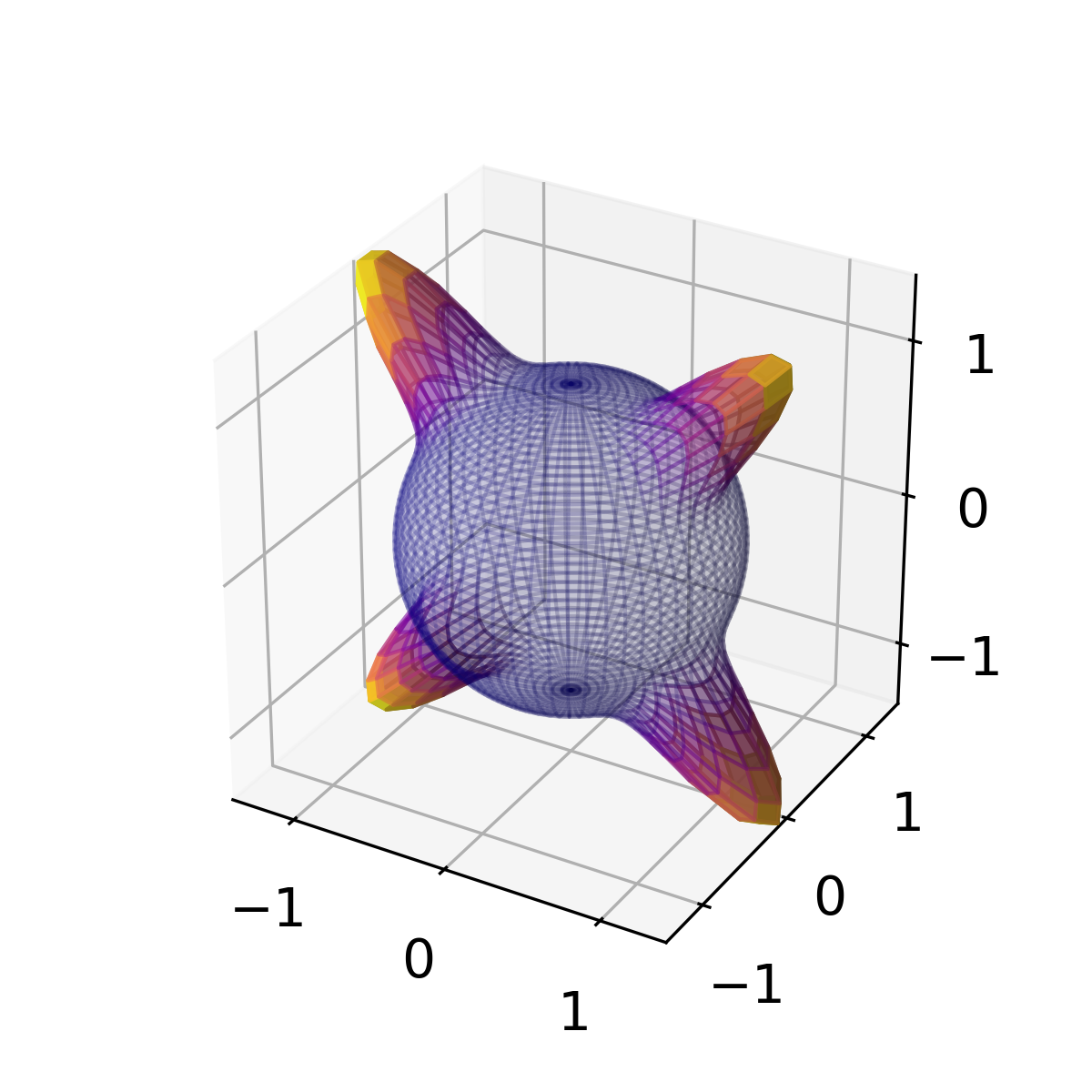}} \\
	\hline
\end{longtable*}

\autoref{tab:enr} presents the rovibrational and hyperfine energies of cluster states for selected values of the total angular momentum quantum number $F$, together with the corresponding hyperfine energy shifts and wavefunction mixing coefficients.
Cluster states with the same total spin-rovibrational symmetry but different nuclear spin components (i.e., \emph{ortho} or \emph{para}) are mixed in the wavefunction together with other spin-rovibrational states of the same total symmetry (see the main manuscript for coupling rules of rovibrational and nuclear spin states).
For example, when the squared absolute value of the coefficient $|c|^2$ is approximately 0.5 for two hyperfine cluster states of B$_2$ symmetry at $F=60$, the wavefunction can be written as $\approx\pm\sqrt{0.5}|\text{A}_1\rangle|I=0\rangle \pm \sqrt{0.5}|\text{B}_2\rangle|I=1\rangle$, where A$_1$ and B$_2$ denote the spatial symmetries of the rovibrational states (see Fig.~3.a of the main manuscript).
With increasing the level of rotational excitation, the hyperfine shifts increase, with cluster states exhibiting stronger \emph{ortho}-\emph{para} mixing.

\begin{table}
\centering
\renewcommand{\arraystretch}{1.2}
\begin{tabular*}{0.5\linewidth}{@{\extracolsep{\fill}}ccccrc}
\hline
$F$ & $\Gamma$ & $E$ (cm$^{-1}$) & $E_\text{rv}$ (cm$^{-1}$) & $\Delta$ (kHz) & $|c|^2$ \\
\hline
40 & B$_1$ &    17749.5716483 &    17749.5716483 &     -1.7 &  1.000 \\
40 & B$_1$ &    17749.5819864 &    17749.5819869 &    -14.3 &  1.000 \\
40 & B$_2$ &    17749.5716470 &    17749.5716476 &    -17.6 &  1.000 \\
40 & B$_2$ &    17749.5819869 &    17749.5819869 &      1.7 &  1.000 \\
45 & B$_1$ &    21121.9041702 &    21121.9041742 &   -119.0 &  0.984 \\
45 & B$_1$ &    21121.9043908 &    21121.9043873 &    103.1 &  0.984 \\
45 & B$_2$ &    21121.9041709 &    21121.9041743 &   -103.6 &  0.984 \\
45 & B$_2$ &    21121.9043903 &    21121.9043873 &     87.8 &  0.984 \\
50 & B$_1$ &    24708.8187665 &    24708.8187749 &   -251.3 &  0.930 \\
50 & B$_1$ &    24708.8188870 &    24708.8188791 &    235.5 &  0.930 \\
50 & B$_2$ &    24708.8187660 &    24708.8187749 &   -264.9 &  0.932 \\
50 & B$_2$ &    24708.8188875 &    24708.8188791 &    249.1 &  0.932 \\
55 & B$_1$ &    28457.3453807 &    28457.3454107 &   -898.1 &  0.571 \\
55 & B$_1$ &    28457.3454493 &    28457.3454198 &    882.3 &  0.571 \\
55 & B$_2$ &    28457.3453808 &    28457.3454107 &   -895.8 &  0.563 \\
55 & B$_2$ &    28457.3454492 &    28457.3454199 &    880.1 &  0.563 \\
56 & B$_1$ &    29221.4694136 &    29221.4694476 &  -1020.4 &  0.508 \\
56 & B$_1$ &    29221.4694828 &    29221.4694492 &   1004.7 &  0.508 \\
56 & B$_2$ &    29221.4694136 &    29221.4694476 &  -1020.7 &  0.515 \\
56 & B$_2$ &    29221.4694828 &    29221.4694493 &   1005.0 &  0.515 \\
57 & B$_1$ &    29989.2832299 &    29989.2832652 &  -1059.1 &  0.506 \\
57 & B$_1$ &    29989.2833003 &    29989.2832655 &   1043.4 &  0.506 \\
57 & B$_2$ &    29989.2832299 &    29989.2832655 &  -1068.7 &  0.501 \\
57 & B$_2$ &    29989.2833003 &    29989.2832652 &   1052.9 &  0.501 \\
58 & B$_1$ &    30760.2866593 &    30760.2866954 &  -1084.4 &  0.503 \\
58 & B$_1$ &    30760.2867310 &    30760.2866954 &   1068.6 &  0.503 \\
58 & B$_2$ &    30760.2866593 &    30760.2866954 &  -1082.2 &  0.504 \\
58 & B$_2$ &    30760.2867310 &    30760.2866954 &   1066.5 &  0.504 \\
59 & B$_1$ &    31533.9665671 &    31533.9666038 &  -1102.7 &  0.504 \\
59 & B$_1$ &    31533.9666401 &    31533.9666038 &   1087.0 &  0.504 \\
59 & B$_2$ &    31533.9665671 &    31533.9666039 &  -1102.8 &  0.503 \\
59 & B$_2$ &    31533.9666401 &    31533.9666038 &   1087.1 &  0.503 \\
60 & B$_1$ &    32309.7952719 &    32309.7953093 &  -1121.2 &  0.504 \\
60 & B$_1$ &    32309.7953462 &    32309.7953093 &   1105.5 &  0.504 \\
60 & B$_2$ &    32309.7952719 &    32309.7953093 &  -1122.6 &  0.503 \\
60 & B$_2$ &    32309.7953462 &    32309.7953093 &   1106.9 &  0.503 \\
\hline
\end{tabular*}
\caption{Cluster state rovibrational energies ($E_\text{rv}$), hyperfine energies ($E$), and hyperfine shifts ($\Delta=E-E_\text{rv}$).
Also shown are the squared absolute values of the leading coefficients $|c|^2$ in the hyperfine wavefunction.
The results are listed for the ground vibrational state and selected values of the total angular momentum quantum number $F$.
The state symmetry $\Gamma$ represent the symmetry of the product of spatial rovibrational and nuclear spin functions.
}\label{tab:enr}
\end{table}

\begin{figure*}[b!]
\centering
\includegraphics[width=\linewidth]{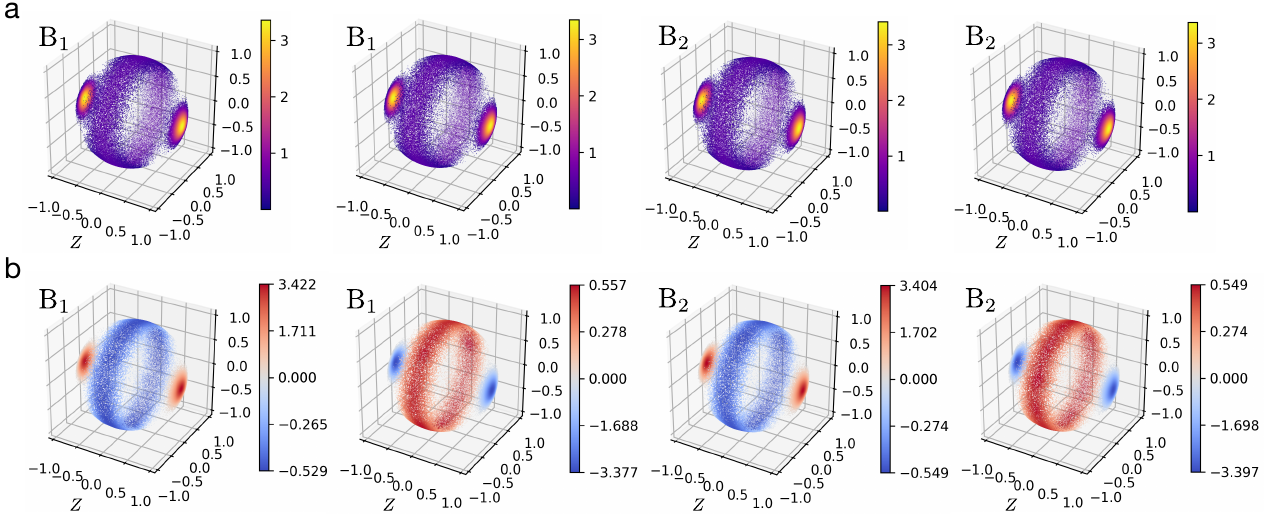}
\caption{
  (a) Rotational probability density distributions of hydrogen nuclei in the laboratory frame, plotted in Cartesian coordinates  on a unit sphere, for four hyperfine cluster states of H$_2$S at $F=60$ and $m_F=60$.
  The plots for different states are presented in the same order as they are listed in \autoref{tab:enr}.
  (b) Corresponding laboratory-frame $Z$-components of spin-densities.
}\label{fig:dens}
\end{figure*}

In \autoref{fig:dens}.a, the probability density distributions of hydrogen nuclei in the laboratory frame are plotted for four hyperfine cluster states at $F=60$ and $m_F=60$ (as listed in \autoref{tab:enr}).
\autoref{fig:dens}.b presents the corresponding spin-densities for the same states.
The rotational densities illustrate the spatial localization of the two hydrogen nuclei: one nucleus follows a larger orbital trajectory, while the other remains localized along the axis perpendicular to the rotation of the first hydrogen.
Classically, this corresponds to molecular rotation about an axis aligned with one of the S--H bonds.

The strong \emph{ortho}-\emph{para} mixing in hyperfine cluster states gives rise to nuclear spin polarization along the rotational axis, with the two hydrogen nuclei exhibiting opposite spin orientations (due to the Pauli exclusion principle).
States of the same symmetry, coupled via hyperfine spin-rotation interaction, exhibit opposite relative orientation of spin polarization vectors.
The near-degenerate pairs of states with B$_1$ and B$_2$ symmetry exhibit the same spin polarization, but reverse their spin projections from one pair of degenerate states to another.

\section{Stark effect in hyperfine cluster states}

\begin{figure}
\centering
\includegraphics[width=0.4\linewidth]{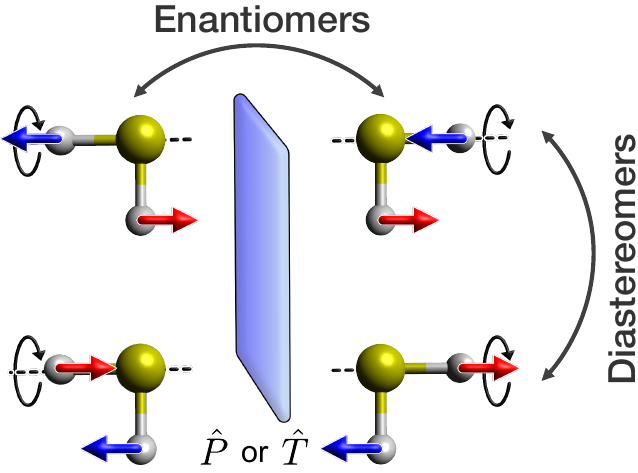}
\caption{
  Schematic representation of spin-induced rotating diastereomers, which exhibit different energies and opposite orientations of spin polarization vectors along the rotational axis.
}\label{fig:diastereomers}
\end{figure}

An electric field applied along the axis of rotation produces the effect of deracemization of a mixture of rotating enantiomers (see Eq.~1 and Fig.~2.a of the main manuscript).
The positive or negative direction of the applied field relative to the rotational angular momentum, determines the predominant enantiomer~\cite{Owens_PRL121_2018}.
The same effect can be achieved by reversing the direction of rotation (i.e., changing $m_F$ to $-m_F$).

When considering hyperfine interactions with nuclear spins, the initially degenerate rotating enantiomers, corresponding to rotations about the S--H$_1$ and S--H$_2$ bonds, form linear combinations and become energetically split.
This splitting results in a set of diastereomers, which are distinguished not only by their energy but also by the opposite orientations of their nuclear spin polarization vectors, as illustrated in \autoref{fig:diastereomers}.

\begin{figure*}
\centering
\includegraphics[width=0.8\linewidth]{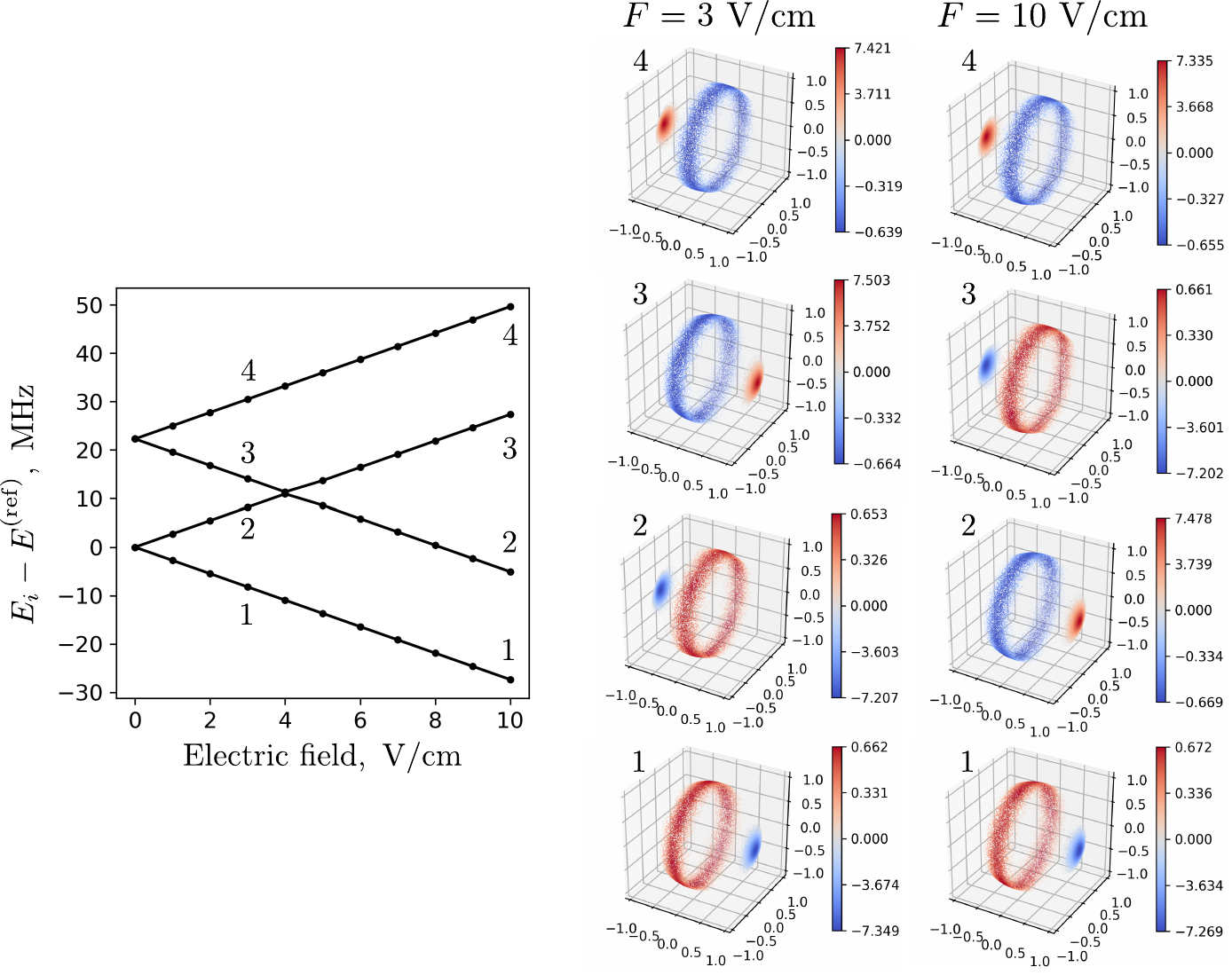}
\caption{
  Stark energies for hyperfine cluster states at $F=60$ and $m_F=60$, plotted relative to the reference energy $E_1^\text{(ref)}=32309.7952719$~cm$^{-1}$ (energy of the lowest hyperfine cluster state).
  Spin-polarization distributions are shown for four Stark-split cluster states at field strengths of 3~V/cm and 10~V/cm.
}\label{fig:stark}
\end{figure*}

This schematic representation of spin-induced diastereomers is supported by simulations of the Stark effect.
The Stark effect, computed for an external electric field applied along the laboratory $Z$-axis for cluster states at $F=60$ and $m_F=60$, is shown in \autoref{fig:stark}.
Additionally, spin-density distributions at different field strengths are plotted to illustrate differences in spin polarization between the two diastereomers.

\end{document}